\documentclass[conference]{IEEEtran}
\IEEEoverridecommandlockouts
\usepackage{cite}
\usepackage{amsmath,amssymb,amsfonts}
\usepackage{algorithmic}
\usepackage{graphicx}
\usepackage{subcaption}
\usepackage{textcomp}
\usepackage{xcolor}
\usepackage{pifont} 
\usepackage{multirow}
\usepackage{rotating}
\usepackage[table]{xcolor}
\newcommand{\cmark}{\checkmark}
\newcommand{\xmark}{\ding{55}}
\usepackage{threeparttable}
\usepackage{subcaption}
\def\BibTeX{{\rm B\kern-.05em{\sc i\kern-.025em b}\kern-.08em
    T\kern-.1667em\lower.7ex\hbox{E}\kern-.125emX}}

\usepackage{pifont} 

\usepackage[ruled,vlined,linesnumbered]{algorithm2e}
\usepackage{wrapfig}

\let\oldnl\nl
\newcommand{\nonl}{\renewcommand{\nl}{\let\nl\oldnl}}

\usepackage[font=small,skip=3pt]{caption}
\usepackage{subcaption}
\usepackage{booktabs} 
\usepackage{bbm}
\usepackage{amsthm} 
\usepackage{amsmath}

\SetKwInput{KwData}{Input}
\SetKwInput{KwResult}{Output}

\usepackage{mathtools}

\usepackage{enumitem}
\usepackage{setspace}
\DontPrintSemicolon

\usepackage{amssymb}
\renewcommand{\cmark}{\checkmark}
\renewcommand{\xmark}{\text{\sffamily X}}
\usepackage{makecell}

\newcommand{\ourds}{\textit{ESCHER}\xspace}

\begin{document}

\title{\LARGE \bf ESCHER: Efficient and Scalable Hypergraph Evolution Representation with Application to Triad Counting\\
\large (to be published in the 2026 IEEE International Parallel and Distributed Processing Symposium (IPDPS))}

\author{
\IEEEauthorblockN{S.~M.~Shovan\IEEEauthorrefmark{3}\textsuperscript{*},
Arindam Khanda\IEEEauthorrefmark{3}\textsuperscript{*},
Sanjukta Bhowmick\IEEEauthorrefmark{2},
Sajal K.~Das\IEEEauthorrefmark{3}}
\IEEEauthorblockA{\IEEEauthorrefmark{3}Missouri University of Science and Technology, Rolla, MO, USA\\
Email: \{sskg8, akkcm, sdas\}@mst.edu}
\IEEEauthorblockA{\IEEEauthorrefmark{2}University of North Texas, Denton, TX, USA\\
Email: Sanjukta.Bhowmick@unt.edu}
\thanks{\textsuperscript{*}S.~M.~Shovan and Arindam Khanda contributed equally to this work.}
}

\maketitle

\begin{abstract}
Higher-order interactions beyond pairwise relationships in large complex networks are often modeled as hypergraphs. Analyzing hypergraph properties such as triad counts is essential, as hypergraphs can reveal intricate group interaction patterns that conventional graphs fail to capture. In real-world scenarios, these networks are often large and dynamic, introducing significant computational challenges. Due to the absence of specialized software packages and data structures, the analysis of large dynamic hypergraphs remains largely unexplored. Motivated by this gap, we propose \textsc{ESCHER}, a GPU-centric parallel data structure for \textbf{E}fficient and \textbf{Sc}alable \textbf{H}ypergraph \textbf{E}volution \textbf{R}epresentation, designed to manage large-scale hypergraph dynamics efficiently. 
We also design a hypergraph triad-count update framework that minimizes redundant computation while fully leveraging the capabilities of \textsc{ESCHER} for dynamic operations. 
We validate the efficacy of our approach across multiple categories of hypergraph triad counting, including hyperedge-based, incident-vertex-based, and temporal triads.
Empirical results on both large real-world and synthetic datasets demonstrate that our proposed method outperforms existing state-of-the-art methods, achieving speedups of up to $104.5\times$, $473.7\times$, and $112.5\times$ for hyperedge-based, incident-vertex-based, and temporal triad types, respectively.

\end{abstract}


\section{Introduction}
\pagestyle{plain} 
\pagenumbering{arabic}

Graphs are powerful mathematical tools for representing complex systems of interacting entities. The nodes represent the entities and the edges, the dyadic interactions between them. Despite their use in diverse disciplines from bioinformatics to cybersecurity, from smart city planning to drug design, graph models are limited in that they can only represent pairwise interactions. Many systems exhibit polyadic reactions that cannot be easily decomposed into pairwise forms.  Chemical reactions between three elements do not imply that any pair of these elements can also react. Similarly, articles with multiple co-authors do not necessarily imply that each pair of authors also have a publication together. Such polyadic reactions are more accurately represented by hypergraphs.

\begin{figure}[h!]
\vspace{-0.15in}
    \centering
    \begin{subfigure}[b]{0.52\linewidth}
        \centering
        \includegraphics[width=0.85\linewidth]{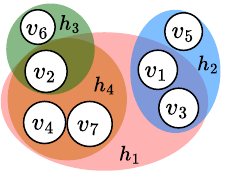}
        \vspace{-0.1in}
        \caption{Hypergraph with 4 hyperedges}
        \label{fig:hypergraph}
    \end{subfigure}
    \hfill
    \begin{subfigure}[b]{0.4\linewidth}
        \centering
        \includegraphics[width=0.85\linewidth]{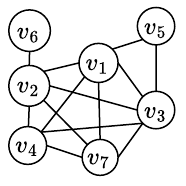}
        \vspace{-0.1in}
        \caption{Equivalent Clique Graph}
        \label{fig:clique}
    \end{subfigure}
    \caption{An example of a hypergraph (a) and its equivalent graph representation (b).In the hypergraph, the colored circles ($h1$, $h2$, $h3$, and $h4$) represent the hyperedges, and the white circles ($v1$-$v7$) represent the edges. Note that information about hyperedge $h4$ is lost in the graph representation, as it is a subset of $h1$}
    \label{fig:example}
\end{figure}

Hypergraphs are generalization of graphs, where each hyperedge is defined by a set of nodes, not just two. Fig~\ref{fig:hypergraph} shows a hypergraph with four edges and seven vertices. Fig~\ref{fig:clique} shows the corresponding graph representation, when the hyperedges are decomposed to dyadic edges. Note that information on hyperedge $h_4$, whose interactions are a subset of hyperedge $h_1$ is lost in the graph representation.

While there exist several effective data structures (adjacency lists,  sparse matrix formats, such as CSR) to implement graphs, developing efficient data structures for hypergraphs is yet an ongoing problem. The primary challenge arises due to 
the variable (rather than fixed) number of vertices per hyperedge, which does not immediately translate into the existing graph formats. The problem is further exacerbated as the system  becomes large (leading to high cardinality hyperedges) and/or dynamic (where vertices and hyperedges can be added or deleted). Further, hypergraph algorithms require mapping from vertex to hyperedge and vice versa. Most data structures cannot efficiently accommodate this two-way mapping. Finally, while hypergraph data structures can support graph algorithms, the process is significantly slower. This presents an insidious but important problem in that there exists no system to provide an equitable comparison between graph and hypergraph models. Such comparisons are essential to study the relative merits of the models.

\begin{figure*}[!t]
\centering
\begin{subfigure}{0.36\textwidth}
    \includegraphics[width=\linewidth]{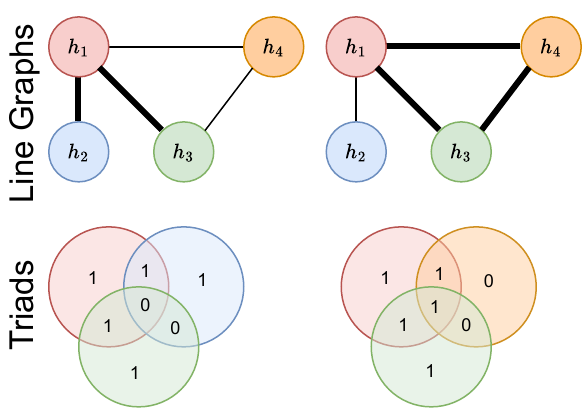}
    \caption{Hyperedge-based Triads}
    \label{fig:hyppro}
\end{subfigure}\hfill
\begin{subfigure}{0.42\textwidth}
    \includegraphics[width=\linewidth]{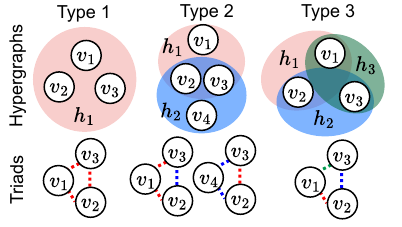}
    \caption{Incident Vertex-based Triads}
    \label{fig:types}
\end{subfigure}\hfill
\begin{subfigure}{0.22\textwidth}
    \includegraphics[width=\linewidth]{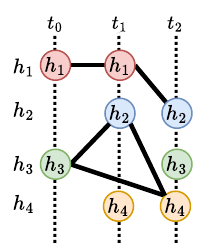}
    \caption{Temporal Hypergraph Triads}
    \label{fig:tempHyp}
\end{subfigure}
\caption{ Hypergraph Triads.(a) Hyperedge-based; (b) Types of Incident Vertex based with different hyperedge overlaps; (c) Temporal Triads.}
\label{fig:all_triads}
\vspace{-0.2in}
\end{figure*}

 We address these issues by designing and developing {\bf ESCHER} ({\bf E}fficient and {\bf Sc}alable {\bf H}ypergraph {\bf E}volution {\bf R}epresentation), to the best of our knowledge, {\em the first GPU-based data structure for large dynamic hypergraphs}. ESCHER supports both hyperedge insertions and deletions, as well as incident vertex modifications, and handles two-way dynamics, thereby enabling efficient updates in both directions. ESCHER addresses the variable hyperedge size problem by pre-allocating GPU memory through batch processing, partial memory allocation, and memory block reuse in overflow cases. ESCHER leverages balanced complete binary trees (CBTs) to optimize memory usage while minimizing computational overhead for insertions and deletions.

We demonstrate the efficacy of ESCHER through its application to computing triads. Triads in hypergraphs are analogous to triangles in graphs. Similar to the importance of triangles, hypergraph triads have application in a wide range of domains, including community detection in social networks ~\cite{ke2019community}, gene co-relations in biological networks~\cite{kong2019hypergraph}, and recommendation systems \cite{wei2022dynamic}.  Unlike triangles, hypergraph triads can be of many types (see Figure~\ref{fig:all_triads}), from triads of hyperedges, to triads of vertices, to dynamic and temporal triads. An efficient data structure should be able to handle all these variations, including the special case of triangles in graphs, as well as exhibit high scalability and performance.

We compare ESCHER with different static and dynamic triad and triangle computing algorithms (see Table~\ref{tab:method-comparison}) and demonstrate that ESCHER indeed provides a competitive, and in many cases improved performance. To summarize, {\bf our main contributions} are:
\begin{itemize}
\item Design and develop ESCHER, a novel GPU-based data structure for dynamic hypergraphs.

\item Develop a parallel framework to update different types of hypergraph triad counts efficiently by leveraging the capabilities of \ourds.

\item Demonstrate ESCHER can count dynamic hyperedge-based triads on average $37.8\times$ faster than the parallel static method, MoChY.

\item Demonstrate ESCHER can count temporal hyperedge-based triads  on average $36.3\times$ faster than a parallel competitive temporal algorithm, THyMe+.

\item Demonstrate ESCHER can count triangles in dynamic graphs competitive to  Hornet, a parallel dynamic graph data structure. 

\item Demonstrate ESCHER can compute different types of vertex-based triads in parallel. To the best of our knowledge, there is no parallel dynamic method for such triads.
\end {itemize}

Together, our experiments show that ESCHER is a scalable and efficient data structure that can be universally applied to dynamic hypergraph and graph analytics.




\section{Definitions}
\label{sec:Preliminaries}


A \textit{hypergraph} is a generalization of a graph, where an edge, called a \textit{hyperedge} is a subset of vertices representing a group interaction\cite{ccatalyurek2001fine}.
Let $G(V, E)$ be an undirected hypergraph, where $V$ is the vertex set and $E = \{h_1, \dots, h_{|E|}\}$ is the hyperedge set, where $h_i \subseteq V$ represents the $i$-th hyperedge. 

If a vertex $v$ belongs to hyperedge $h_i$, the vertex is \textit{incident} to $h$. 
Considering $E_v = \{ h_i \in E : v \in h_i \}$ is the set of hyperedges that contain $v$,  the \textit{degree} of a vertex $ v \in V$ is defined as $\deg(v) = |E_v|$.
The degree or \textit{cardinality} of a hyperedge $h_i$, denoted by $|h_i|$, is the number of vertices it contains: $|h_i| = |\{v \in V : v \in h_i\}|$. 





\begin{figure*}[!t]
\centering
\begin{subfigure}{\textwidth}
    \includegraphics[width=\linewidth]{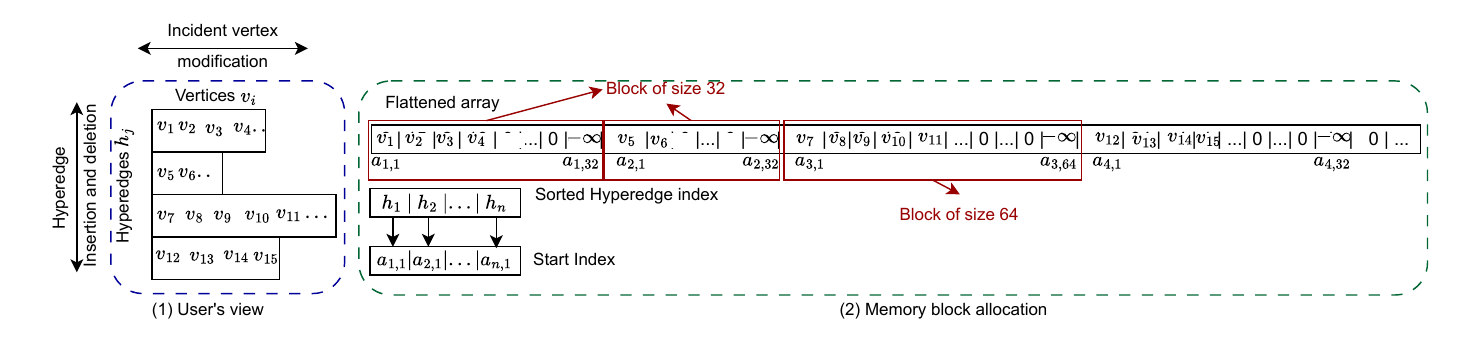}
    \vspace{-0.3in}
    \caption{Hyperedge-to-incident-vertex (h2v) mapping representation.}
    \label{fig:ds_1}
\end{subfigure}\hfill
\begin{subfigure}{\textwidth}
    \includegraphics[width=\linewidth]{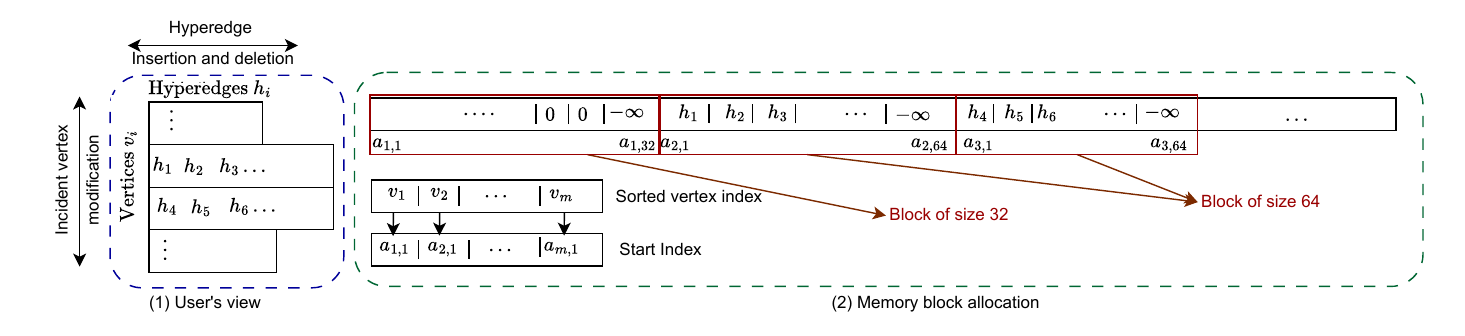}
    \vspace{-0.3in}
    \caption{Vertex-to-incident-hyperedge (v2h) mapping representation.}
    \label{fig:ds_2}
\end{subfigure}
\caption{Hypergraph stored in flattened memory blocks in GPU.}
\label{fig:ds_overview}
\vspace{-0.15in}
\end{figure*}

{\em Hypergraph Triads.} A hypergraph triad represents three connected hyperedges or incident vertices. Hypergraph triads can be classified as follows (also illustrated in Figure~\ref{fig:all_triads});
\begin{itemize}
\item {\em Hyperedge Triads:} Let $ \mathbb{H}((h_i, h_j, h_k)) $ be a function that considers each pair of neighboring hyperedges $ (h_i, h_j) \in \binom{N_{h_i}}{2} $ and $h_k \in N_{h_i} \cup N_{h_j}$ to count different types of triadic motifs in a hypergraph.
Different triadic patterns involving hyperedges $h_i$, $h_j$, and $h_k$ can be represented using Venn diagrams, resulting in $2^7$ possible configurations. After removing symmetric triads, 26 unique hypergraph triads can be formed~\cite{lee2020hypergraph}.

Fig.~\ref{fig:hyppro} illustrates two triad formations from Fig.~\ref{fig:hypergraph} in a \textit{line graph}, a representation where hyperedges are treated as nodes, with an edge between two nodes if their corresponding hyperedges share at least one vertex\cite{liu2022high}. The corresponding Venn diagrams of the line graphs show whether common vertices exist among the hyperedges. 

\item{\em Incident Vertex Triads:} These triads represent connections between three incident vertices and are categorized based on the membership of the vertices in hyperedges.
Fig.~\ref{fig:all_triads}(b) illustrates three categories: Type~1 (all 3 pairs of vertices belong to same hyperedge), Type~2 ( a subset of vertex pairs (1 or 2) to same hyperedge), and Type~3 (all 3 vertex pairs belong to different hyperedges),  as defined in~\cite{bhattacharya2025statistical}. In these triads, each dotted edge indicates a common hyperedge membership of two endpoints.

\item{\em Dynamic and Temporal Triads:} Dynamic triads are triads in dynamic hypergraphs that change as vertices and hyperedges are added or deleted. Temporal triads are a subclass of dynamic triads, where the hyperedges are associated with time stamps. In such triads, both the connections and their order in time matter. Here, only triads formed over a given time window are considered. For example, For three connected temporal hyperedges $h_i, h_j,$ and $h_k$ with arrival times $t_i, t_j,$ and $t_k$ satisfying $t_i < t_j < t_k$, they form a valid temporal triad only if $t_k - t_i \leq t_\delta$, where $t_\delta$ is a predefined time window.
\end{itemize}

Table~\ref{tab:related_works:diff_methods} lists the state-of-the-art software for computing triads in dynamic hypergraphs and graphs. More details about the software are provided in Section~\ref{related-work}.

While many parallel algorithms~\cite{sahu2024shared, haryan2022shared, khanda2023parallel,shovan2025parallel} and data structures~\cite{busato2018hornet, green2016custinger} exist for analyzing dynamic graphs, fewer research efforts have focused on handling dynamic hypergraphs. Through ESCHER, we aim to provide a GPU-based parallel data structure for analyzing large dynamic hypergraphs.

\begin{table}[ht]
\centering
\vspace{-0.07in}
\small
\caption{Methods supporting different triad types.}
\label{tab:related_works:diff_methods}
\setlength{\tabcolsep}{4pt} 
\renewcommand{\arraystretch}{1.1} 
\begin{tabular}{|l|l|c|c|c|c|}
\hline
\textbf{Method} & \textbf{Triad Types} &
\makecell{\rotatebox{90}{\textbf{Hypergraph}}} &
\makecell{\rotatebox{90}{\textbf{Dynamic}}} &
\makecell{\rotatebox{90}{\textbf{Temporal}}} &
\makecell{\rotatebox{90}{\textbf{Parallel}}} \\
\hline
MoCHy~\cite{lee2020hypergraph} & Hyperedge-based       & \cmark & \xmark & \xmark & \cmark \\
THyMe+~\cite{lee2021thyme+}    & Hyperedge-based       & \cmark & \xmark & \cmark & \xmark \\
StatHyper~\cite{bhattacharya2025statistical} & Incident vertex-based & \cmark & \xmark & \xmark & \xmark \\
Hornet~\cite{busato2018hornet} & Vertex-based          & \xmark & \cmark & \xmark & \cmark \\
\textbf{\ourds (our)}                           & All                   & \cmark & \cmark & \cmark & \cmark \\
\hline
\end{tabular}
\label{tab:method-comparison}
\vspace{-0.05in}
\end{table}

\section{Description of ESCHER}
\label{sec:proposed}

We now describe the design and operations supported by \ourds  to support large-scale hypergraph dynamics on GPU. 

{\em Enabling Multiple Formats.} Hypergraphs can be represented in different formats, including
 bipartite graphs, line graphs, clique graphs, and $s$-clique graphs. Unlike graphs, it is not always possible to translate from one format to another, as some information may be lost in the translation. Each format has its benefits. Algorithms are designed specifically for a given format to utilise that that particular information the format encodes. Thus, an efficient hypergraph data structure should support all these formats.

 Hypergraphs can be represented as either sets of hyperedges, each containing sets of vertices (denoted here as {\em h2v}) or sets of vertices belonging to sets of hyperedges (denoted here as {\em v2h}). Line graphs represent connections between hyperedges (denoted here as {\em h2vh}). By supporting these three mappings, all different formats can be implemented. For example, h2v supports the bipartite representation, v2h supports clique graphs, and h2h is a direct representation of the line-graph.

  \ourds offers a single schema that supports all three mappings, with the user's view being similar to an adjacency list. In the mapping h2v, each hyperedge in the list maintains the list of vertices belonging to it, Fig.~\ref{fig:ds_1}; In the mapping v2h, each vertex in the list maintains the list of hyperedges to which it belongs, Fig.~\ref{fig:ds_2} In the mapping h2h (not shown in Figure), each hyperedge in the list, maintains the list of hyperedges to which it is connected. Note that the mapping v2v, which represents dyadic graphs, can also be accommodated through this schema. For the remainder of this section, we will describe the data structure and operations supported by \ourds through the lens of h2v mapping. The same data structure and operations are used for the other mappings as well. Table~\ref{tab:mappings} provides an overview of the supported operations and their effects for each mapping.


\begin{table}[h]
\centering
\caption{Supported operations in different mappings.}
\small 
\begin{tabular}{|p{0.5cm}|p{2.6cm}|p{1.8cm}|p{2.2cm}|}
\hline
\textbf{Map.} & \textbf{User's view} & \textbf{Vertical Ops.} & \textbf{Horizontal Ops.} \\
\hline
h2v & incident vertex list per hyperedge & hyperedge ins./del. & incident vertex ins./del. \\ \hline
v2h & incident hyperedge list per vertex & vertex ins./del. & incident hyperedge ins./del. \\ \hline
h2h & neighboring hyperedge list per hyperedge & hyperedge ins./del. & hyperedge neighbor ins./del. \\
\hline
\end{tabular}
\label{tab:mappings}
\\\textbf{Map.}: Mapping or representation, \textbf{Ops.}: Operations 
\vspace{-0.15in}
\end{table}


\subsection{Data Structures of \ourds}


\ourds consists of three main components: \textit{incident list view}, \textit{memory blocks}, and a \textit{block manager}.

\textbf{Incident List.} 
This is an abstraction of the internal workings of the data structure and can be viewed as a list of lists. Fig.~\ref{fig:ds_overview}(a) shows an h2v incident list. The local ID of hyperedge $h_j$ is displayed vertically alongside its list of vertices.

The updates to h2v mapping can be of two types; First is the addition or deletion of vertices to the existing hyperedges. This changes the horizontal lists, so we term this as {\em horizontal update}. Second is the addition or deletion of hyperedges. This changes the vertical lists, so we term this as {\em vertical update}. Both types of updates are supported by \ourds.



\textbf{Memory Block.}
Dynamic memory allocation is challenging in GPUs; thus, maintaining a dynamic list of lists is inefficient. We address this issue by flattening the vertex lists of all hyperedges into a large preallocated 1-D memory array (denoted as $\mathbb{A}$).
To align with the GPU’s warp size and optimize processing, each hyperedge $h_j$ with cardinality $d_j$ is assigned a memory block of size $\lceil (d_j + 1)/32 \rceil \cdot 32$.
An extra slot at the end of each block is reserved for metadata to support dynamic hyperedge memory management, as described in Section~\ref{subsec:ops_on_ds}.



Fig.~\ref{fig:ds_1}(b) illustrates the memory blocks storing hyperedges. The first block, starting at $a_{1,1}$, corresponds to $h_1$ and occupies 32 units, while the third block, spanning $a_{3,1}$ to $a_{3,64}$, occupies 64 units.
An $-\infty$ marker is stored as \textit{metadata} to indicate the end of each block’s allocated memory.

\textbf{Tree-based block manager:}
Deleting hyperedges leaves gaps in the flattened array $\mathbb{A}$, representing unallocated memory. Consequently, inserting a new hyperedge requires either costly shift operations to compact the occupied memory blocks and locate the unallocated memory chunk, or a linear-time search to locate a suitable unallocated block within $\mathbb{A}$.

\begin{figure}[!hbtp]
	\centering
\vspace{-0.1in}	\includegraphics[width=.94\linewidth]{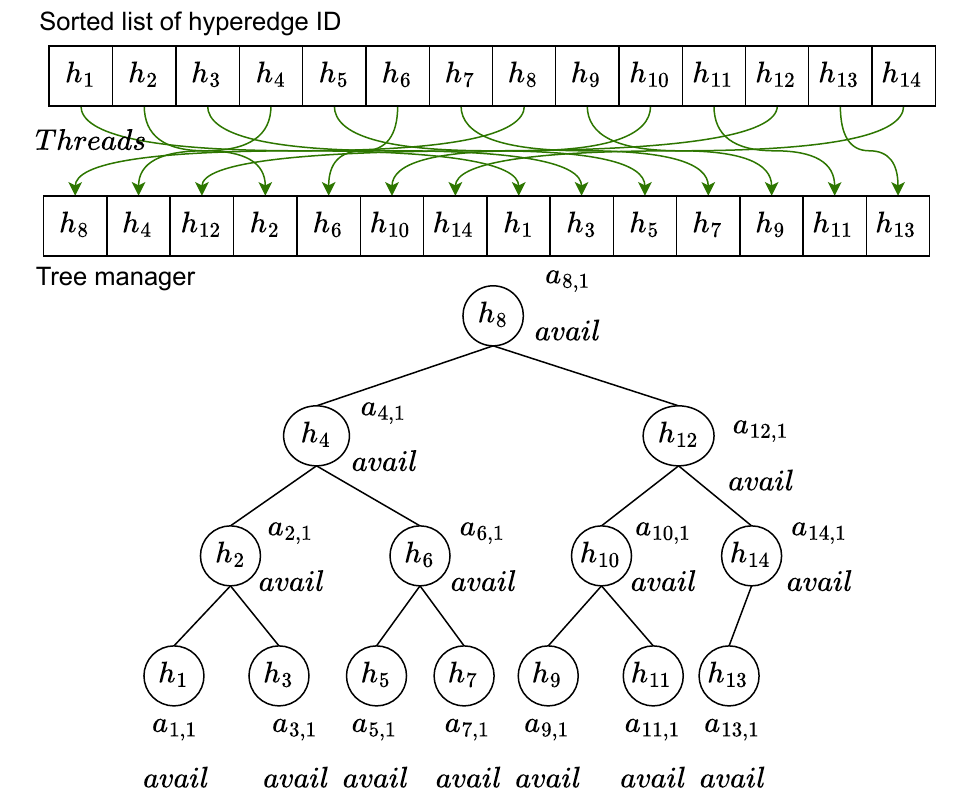}

        \caption{Parallel construction of a complete binary search tree-based block manager. Each node in the tree contains the local ID of a hyperedge $h_j$, starting address of the allocated block $a_{j.1}$, and $avail$ counter storing the count of free nodes in the child subtree.}
        \label{fig:treeConstruction}
    \vspace{-0.15in}
\end{figure}

To facilitate an efficient hyperedge traversal, we design a \textit{complete binary search tree-based} block manager (Fig.~\ref{fig:treeConstruction}). Each node of the tree stores three variables, $h_j$, $a_{j,1}$, and $avail$, denoting the local ID of a hyperedge, the starting address of its memory block, and the number of available blocks in the child subtree. 
The $avail$ counter for each node is initialized to 0 and updated dynamically as the hypergraph evolves (see Section~\ref{subsec:ops_on_ds}). It facilitates parallel hyperedge insertion and block re-allocation.
The block manager supports retrieval of the starting address of any hyperedge $h_j$ in sublinear time and identification of unallocated memory blocks in $O(\log |E|)$ time. Efficiency is maintained by keeping the tree balanced during hyperedge insertions and deletions.

\subsection{Operations on \ourds}
\label{subsec:ops_on_ds}

We now present the operations supported by \ourds.

\textbf{Hypergraph initialization (Parallel construction of tree-based block manager).} 
Given a sorted list of hyperedge IDs, the initialization step constructs the block manager tree by placing each \textit{data item} (hyperedge ID, starting pointer, and $avail$) into the corresponding tree node. Since the hyperedge ids are numbered as consecutive integers, the block manager can be maintained as a \textit{complete binary search tree}. Due to the manager being a complete binary tree, it can be stored as an array--similar to how heaps can be stored as arrays. Each data item is written to the array index that represents the node’s position in the tree.

To optimize this process, parallel threads are assigned to each hyperedge $h_i$ in the sorted list, and they independently compute the position ($idx$) of the associated node in the block manager using Eq.~\ref{eq:1}~\cite{park2001parallel}, where $\ll$ denotes the left shift bit operator, and $tid$ represents the thread ID.
\begin{equation}
\label{eq:1}
   idx = ( (2 \cdot (tid + 1 - (1 \ll \log tid))) + 1 )  \cdot \frac{(1 \ll \log |E|)}{(1 \ll \log tid)}
\end{equation}



Fig.~\ref{fig:treeConstruction} illustrates the parallel construction of the block manager.
After the initial construction, the hyperedge insertion or deletion requires updates to the tree manager. 

\textbf{Searching for Hyperedge.}
The block manager tree helps locate the starting pointer of a hyperedge's memory block efficiently. 
The search operation based on the hyperedge ID follows the standard complete binary search tree-based technique. For a list of $s$ hyperedge IDs, the search takes $O(\frac{s}{p} \cdot \log |E|)$ time, where $p$ is the number of parallel threads and $\log |E|$ is the height of the block manager tree.



\textbf{Deletion of hyperedges (\textit{Vertical Operation}).} 
When a hyperedge is deleted, \ourds searches the corresponding \textit{block manager tree node} by its hyperedge ID and marks it as affected.(Algorithm~\ref{algo:del_new}, Lines~\ref{code:search_start}-\ref{code:search_end}). Only the variable $avail$ of the node is incremented to indicate that the memory block related to the deleted hyperedge is now available. 
A block manager node associated with a deleted hyperedge retains the ID and memory block pointer of the deleted hyperedge. These are only modified or overwritten when the node is reassigned to another hyperedge. Thus, 
 this method avoids marking each element in a hyperedge memory block and simply tracks the deleted block as available for reuse. 
Therefore, after hyperedge deletions, the number of nodes in the block manager tree remains unchanged, containing both existing hyperedges and available memory blocks, and the tree remains sorted and balanced,  without the need to perform expensive tree rotation.

\begin{algorithm}
\caption{Hyperedge deletion}
\DontPrintSemicolon
\small
\label{algo:del_new}
$Affected \gets \emptyset$\;
\tcp{Search and mark affected}
\For{each $h_i$ in deleted hyperedges $Del$ in parallel}{

    $node \gets $ root of $blockManager$\;
    \While{$True$}
    {   \label{code:search_start}
  
        \If{$node < h_i $}
        {

            $node \gets $right child\tcp{right subtree.}
        }
        \ElseIf{$node > h_i$}
        {

            $node \gets $left child\tcp{left subtree.}
        }
        \Else{
             
            Mark the current $node$ as affected by adding it to $Affected$.\;
            Incrementing the node's $avail$ by one.\;

            Break.\;
        }\vspace{-0.03in}

    }\vspace{-0.03in}\label{code:search_end}

}\vspace{-0.03in}
\tcp{Update the $avail$ count}
\While{$Affected$ is not empty}
{
    \For{each node $a \in Affected$ in parallel}{
        Remove the node $a$ from $Affected$\; 
        $parent  = \lfloor(\text{index of $a$}) / 2\rfloor$\;
        Update the $avail$ count of $parent$ node by summing the $avail$ value of its left and right child nodes.\label{code:update_par}\;
        \If{ $parent$ node is not the root}
        {
            Add $parent$ node to $Affected$\;
        }\vspace{-0.03in}
    }\vspace{-0.03in}
}\vspace{-0.03in}

\end{algorithm}

\begin{figure}[!hbtp]
	\centering
\vspace{-0.2in}
    \subfloat[A view of the memory array $\mathbb{A}$ after deleting $h_1,h_5,h_6,h_{10}$.]{%
		\includegraphics[width=0.95\linewidth]{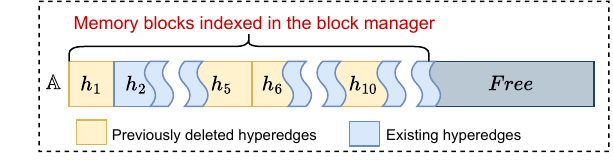}
	}\\\vspace{0.04in}
	\subfloat[Finding $j^{th}$ available node in block manager by thread $Th_j$.]{%
		\includegraphics[width=0.88\linewidth]{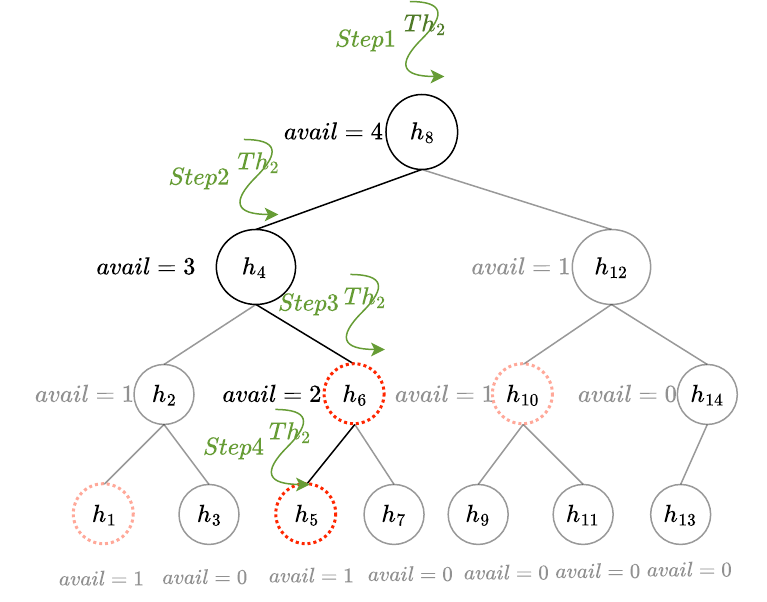}
	}\\ \vspace{0.04in}
    \subfloat[View of $\mathbb{A}$ after case 1 insertion.]{%
		\includegraphics[width=0.95\linewidth]{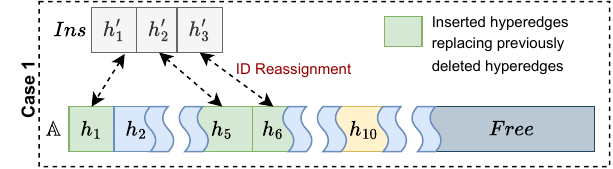}
	}\\ \vspace{0.04in}
    \subfloat[View of $\mathbb{A}$ after case 2 insertion.]{%
		\includegraphics[width=0.95\linewidth]{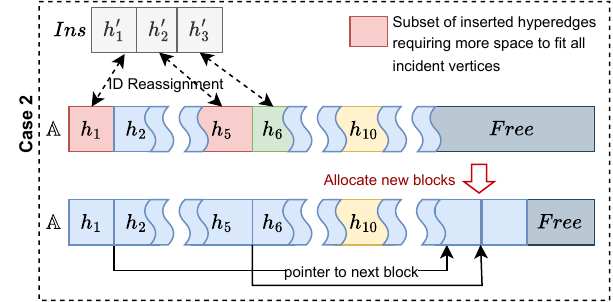}
	}
    \\\vspace{0.04in} 
    \subfloat[View of $\mathbb{A}$ after case 3 insertion.]{%
		\includegraphics[width=0.95\linewidth]{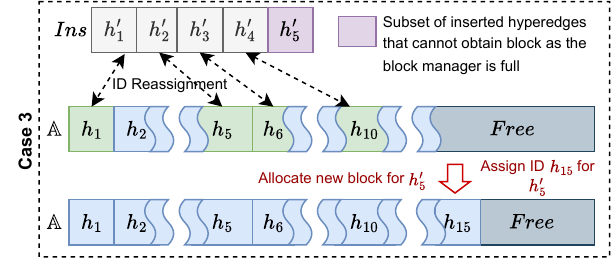}
	}


 \caption{Example illustrating the different cases of deletion and insertion of hyperedges.}
	\label{fig:op}
    \vspace{-0.2in}
\end{figure}

After hyperedge deletion, we update the $avail$ counter at every affected tree node so it reflects the number of available blocks in its child subtree. When a node is marked as affected, Algorithm~\ref{algo:del_new}, Line~\ref{code:update_par} updates its parent by summing the $avail$ values of its left and right children. 
Any node with a modified $avail$ is also marked affected and the update propagates iteratively to the root. The $avail$ value of the root indicates the total available blocks, enabling quick access to available blocks from the block manager root. Fig.~\ref{fig:op} (a) illustrates the deletion of hyperedges.

For a set of deleted hyperedges $Del$, each parallel thread handles one hyperedge from $Del$, marking it as affected in $O(\frac{|Del|}{p} \cdot \log |E| + \log |E|)$ time. Next, the affected nodes are processed in parallel to update their $avail$ values. Since each level of block manager tree requires $O(|Del|)$ work and at most $\log |E|$ levels are needed to propagate updates to the root, this step also runs in $O(\tfrac{|Del|}{p} \cdot \log |E| + \log |E|)$ time.



\textbf{Insertion of Hyperedges (\textit{Vertical Operation}).} 
For each insertion \ourds checks the block manager to see if there is available memory that can be used. Otherwise, the hyperedge is allocated to unindexed memory in the flattened array $\mathbb{A}$. Fig.~\ref{fig:op}(b) illustrates parallel insertion, where each thread starts at the root node $h_8$ and independently searches for its assigned available block using the $avail$ counters. The example highlights thread $th_2$, which traverses the tree and locates the second available block by following Algorithm~\ref{algo:ins_new}.We consider the following three cases for insertion.



\noindent\textit{Case 1. The number of inserted hyperedges can be accomodated in the available blocks.} (Fig.~\ref{fig:op}(c)).


In this case, all inserted hyperedges are accommodated by the available memory blocks indexed in the block manager. The new hyperedge is assugned the same ID, as the previously deleted hyperedge in that block.
. This reassignment avoids tree rebalancing operations that would otherwise be required.

Given a set of inserted hyperedges $Ins$, one thread per hyperedge is launched to identify available memory blocks, via the block manager.
Each thread $Th_j$ is assigned to locate the $j^{th}$ node that contains an available memory block. Because each available node can be reached through a unique path, parallel threads with distinct destinations can proceed asynchronously. Algorithm~\ref{algo:ins_new} guides traversal by comparing the thread ID with $avail$ at each node and selecting an efficient path for each thread. 

While efficient, reassigning new hyperedges to older IDs can lead to misalignment of results. Here we are counting the number of triads and the reassignment does not affect the results. In other applications, we can solve the issue by keeping a map of the new hyperedges to their old IDs.



\begin{algorithm}
\caption{Reassigning available nodes}
\DontPrintSemicolon
\label{algo:ins_new}
\small

\For{each $h_i$ in inserted hyperedges $Ins$ in parallel}{

    $tid \gets $ ID of the assigned thread $Th_j$\;
    $node \gets $ root of $blockManager$\;
    \While{$True $}{
        \label{code:avail_find1}
        \If{$node.avail = tid$ and $node.avail = 1 + $ sum of $avail$ of its left and right children )}{
                Assign the $node$ to $h_j$\;
                $break$\;
            }
        
            \If{$avail$ of left child is less than $tid$}{
                $tid = tid~ -$ left child $avail$ value\;
                Go to the right child node\;                
            }\vspace{-0.03in}
            \Else{
                Go to the left child node\;
            }\vspace{-0.03in}

        }\vspace{-0.03in} 

        \label{code:avail_find2}
    }\vspace{-0.03in}
\end{algorithm}


\noindent\textit{Case 2. The cardinality of the hyperedge is more than any single memory block. } (Fig.~\ref{fig:op}(d)).

A new hyperedge with large cardinality may not be able to fit in a single available block. However, allocating the entire edge to free memory will cause gaps in the array. the assigned block is first filled with incident vertices, and the remaining vertices are placed in a new memory block allocated from the available chunk of the flattened array $\mathbb{A}$. To link consecutive blocks, the metadata (last element) of each block is updated with a pointer to the next block, while the metadata of the final block is set to $-\infty$ to mark the end of the incident vertex list.




\noindent\textit{Case 3. The number of inserted hyperedges is more than the available blocks.} (Fig.~\ref{fig:op}(e)).

When the number of inserted hyperedges exceeds the available blocks in the block manager, the available blocks are first allocated to the new hyperedges, following the same procedure as in Case 1 insertion.
The remaining hyperedges are assigned new memory blocks in parallel from the available memory chunk in $\mathbb{A}$. Since the required block sizes per hypergraph may vary (computed as $\lceil (d_j + 1)/32 \rceil \cdot 32$), a parallel prefix sum is performed 
to determine the starting pointer of blocks assigned to each hyperedge. The hyperedge IDs are then linked to these starting pointers.
The new memory blocks are included in the block manager as new tree nodes via either tree rotation or full reconstruction. Although tree rotation requires less work per level, the large depth in parallel  makes it slower than reconstruction with many GPU threads. Hence, we sort the hyperedges and reconstruct the block manager tree using Eq.~\ref{eq:1}.

With $p$ threads and $|Ins|$ inserted hyperedges, each thread searches for available space in $\log |E|$ steps. The insertion complexity is $O(\frac{|Ins|}{p} \cdot \log |E| \cdot c_{max})$, where $c_{max}$ denotes the maximum cardinality of any hyperedge.

\textbf{Incident vertex insertion/deletion (\textit{Horizontal Operations}):}
For an incident vertex deletion from a hyperedge, the starting address of the associated memory block is retrieved from the block manager, and the block is traversed to locate and remove the vertex. The gap is then filled by shifting subsequent vertex IDs.
On the other hand, for vertex insertion in a hyperedge, the associated memory block is searched for an available position. If space exists, the new vertex ID is added; otherwise, a new block is allocated from the free space of $\mathbb{A}$ and the vertex is inserted there.

For batch vertex modifications on a set of hyperedges, vertices are grouped by hyperedge ID, and a single thread processes each group to avoid race conditions when updating memory blocks. The time complexity for modifying vertices across $\chi$ hyperedges is $O(\frac{\chi}{p} \cdot \log |E| \cdot c_{max})$.



\begin{algorithm}
\caption{Hyperedge-based Triad Count Update}
\DontPrintSemicolon
\label{algo:count_update}
\small
\tcp{Step 1: Find deletion affected hyperedges}
$Aff_{Del} \gets $ deleted hyperedges $Del$\;
\For{each $h_i \in Del$}{
    Add all 1-and 2-hop neighbor hyperedges of $h_i$ to $Aff_{Del}$\;
}

\tcp{Step 2: Find triad count in deletion affected regions}
$count_{Del} \gets $ count unique triads containing hyperedges from $Aff_{Del}$ only\;

\tcp{Step 3: Update the hypergraph}
Delete hyperedges $Del$ in the hypergraph\;
Insert hyperedges $Ins$ in the hypergraph\;

\tcp{Step 4: Find insertion affected hyperedges}
$Aff_{Ins} \gets $ inserted hyperedges $Ins$\;
\For{each $h_i \in Ins$}{
    Add all 1-and 2-hop neighbor hyperedges of $h_i$ to $Aff_{Ins}$\;
}

\tcp{Step 5: Find triad count in all affected regions}
$count_{Ins} \gets $ count unique triads containing hyperedges from $Aff_{Del} \cup Aff_{Ins}$ only\;

\tcp{Step 6: Update the triad count}
$count \gets count - count_{Del} + count_{Ins}$\label{code:update_count}\;

\end{algorithm}

\subsection{Parallel Hypergraph Triad Count Update}
Using \ourds we develop an algorithm for updating hypergraph triads, under hyperedge insertion or deletion. 
Algorithm~\ref{algo:count_update} begins by identifying deletion-affected regions. Step 1 processes deleted hyperedges in parallel and marks them and their 1- and 2-hop neighbors as $Aff_{Del}$. In Step 2, depending on the triad category, any state-of-the-art counting technique is applied to compute the number of triads (denoted $count_{Del}$) formed only by hyperedges in $Aff_{Del}$. Thus, $count_{Del}$ represents the triad count in the deletion-affected regions. Step 3 removes the deleted hyperedges $Del$ and inserts new hyperedges $Ins$. Step 4 identifies insertion-affected regions $Aff_{Ins}$ by processing $Ins$ in parallel. Step 5 applies the same counting technique as in Step 2 to compute triad counts in the combined affected regions $Aff_{Del} \cup Aff_{Ins}$. The updated triad count is obtained by subtracting $count_{Del}$ from the previous total count $count$ and adding the newly formed triads (Step 6, Algorithm~\ref{algo:count_update} Line~\ref{code:update_count}).


Note that \ourds can incorporate any hypergraph computing algorithm, due to supporting multiple formats. Further replacing "hyperedge" with "incident vertex" yields a  supports insertion or deletion of incident vertices and updates the triad count accordingly. Updating triangle counts is also a specialized case of this algorithm. These features highlight the universal applicability of \ourds.

\section{Implementation details}
\label{sec:implementation}
\ourds is implemented in CUDA C++ targeting NVIDIA GPUs. We preallocate extra GPU memory to accommodate hyperedge or incident vertex insertions and thereby avoid unnecessary memory copies between host and device. The preallocation amount can be tuned according to the application. 

\vspace{0.05in}
{\em Hypergraph Maintenance.}
The vertical and horizontal operations on \ourds are implemented as kernels.  Hyperedge deletion requires two kernels. The first, \texttt{markDelete}, identifies hyperedges linked to block manager tree nodes and increments the $avail$ counter in parallel. Then the \texttt{propagateAvail} kernel computes the cumulative number of available blocks. For hyperedge insertion, the first kernel searches for available space in the block manager, the second kernel allocates additional memory blocks, and the last keeps the block manager tree balanced. 
Since insertion Cases 2 and 3 both require new memory blocks, they are handled jointly. Our implementation first marks (i) hyperedges that need extra blocks (Case 2) and (ii) new hyperedges that did not obtain an available block (Case 3). It then allocates memory for all marked hyperedges in one pass. Because all allocations come from the same chunk $\mathbb{A}$ and requested block sizes differ, the starting address of each hyperedge’s block is computed via a parallel prefix sum using CUDA Thrust~\cite{bell2012thrust}.

\vspace{0.05in}

 {\em Hypergraph Triad Count Update.} We implement three types of hypergraph triad counting: (i) hyperedge-based~\cite{lee2020hypergraph}, (ii) incident-vertex-based~\cite{bhattacharya2025statistical}, and (iii) temporal~\cite{lee2021thyme+}. Each triad counting method is implemented as a CUDA kernel (\texttt{countTriads}). As the efficiency of finding triads depends heavily on computing the adjacency-list intersection of two adjacent hyperedges or vertices~\cite{green2014fast}, we optimize it by  parallel sorted set intersection as in \cite{fox2018fast}. 

For dynamic hypergraphs, we implement triad-count updates for different triad types following the framework in Algorithm~\ref{algo:count_update}. Specifically, finding deletion- and insertion-affected subgraphs (Steps 1 and 4) uses two similar kernels that mark the affected hyperedges in parallel and then filter them as the next frontier. Steps 2 and 5 use the \texttt{countTriads} kernel to count triads in the subgraphs. Step 3 updates the hypergraph using the vertical and horizontal operations of \ourds. Step 6 performs a straightforward update of the numeric count.





\begin{figure*}[!hbtp]
	\centering
 
	\subfloat[\scriptsize Effect of varying number of hyperedge changes \label{fig:subfig1}]{%
		\includegraphics[width=0.25\linewidth]{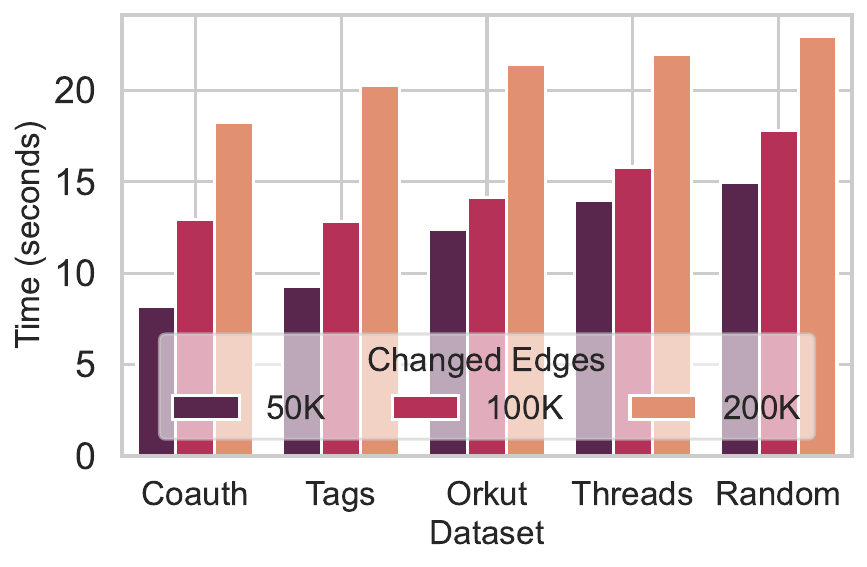}
	}
    \subfloat[\scriptsize  Varying the size of hyperedge counts with 50k fixed changed edges \label{fig:varying_graph_size}]{%
		\includegraphics[width=0.25\linewidth]{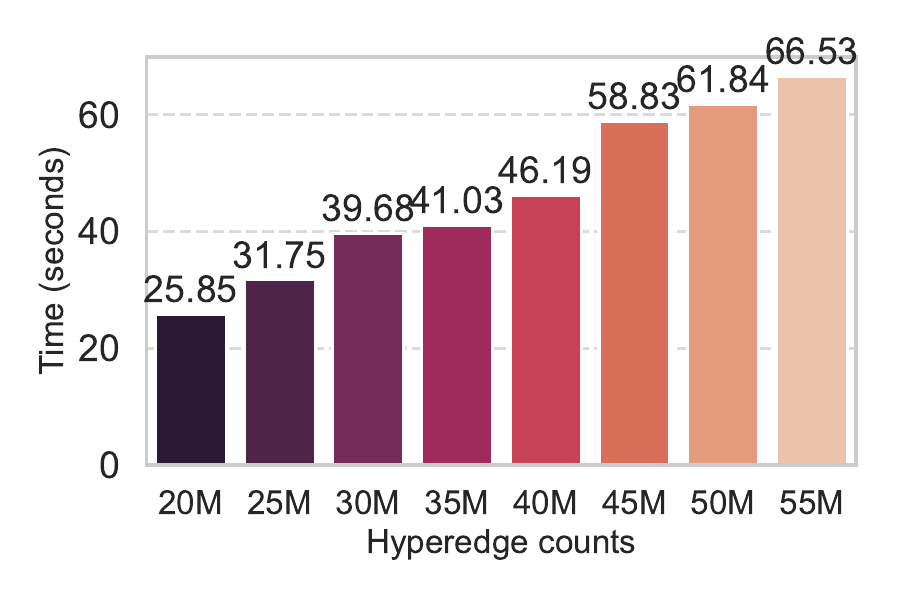}
	}
        \subfloat[ \scriptsize  Effect of varying hyperedge cardinality \label{fig:subfig3}]{%
		\includegraphics[width=0.25\linewidth]{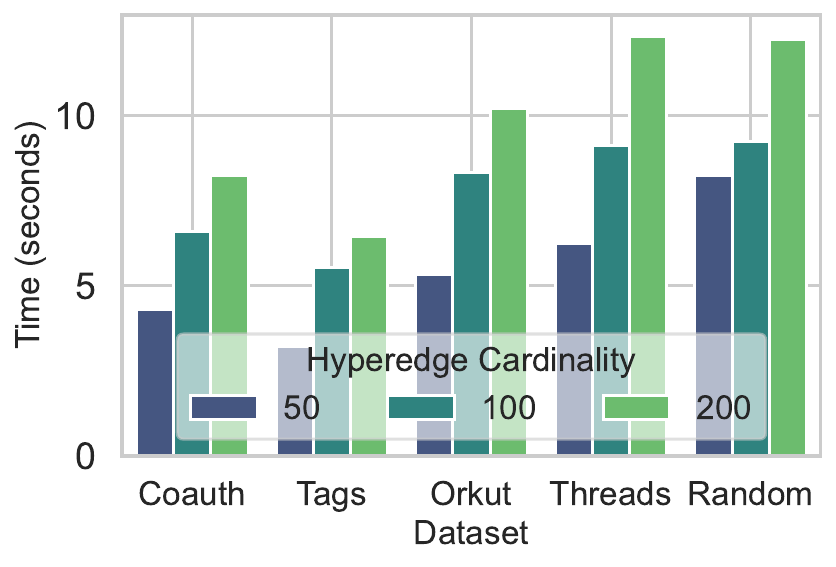}
	}
        \subfloat[ \scriptsize Effect of incident vertex modification\label{fig:subfig4}]{%
		\includegraphics[width=0.25\linewidth]{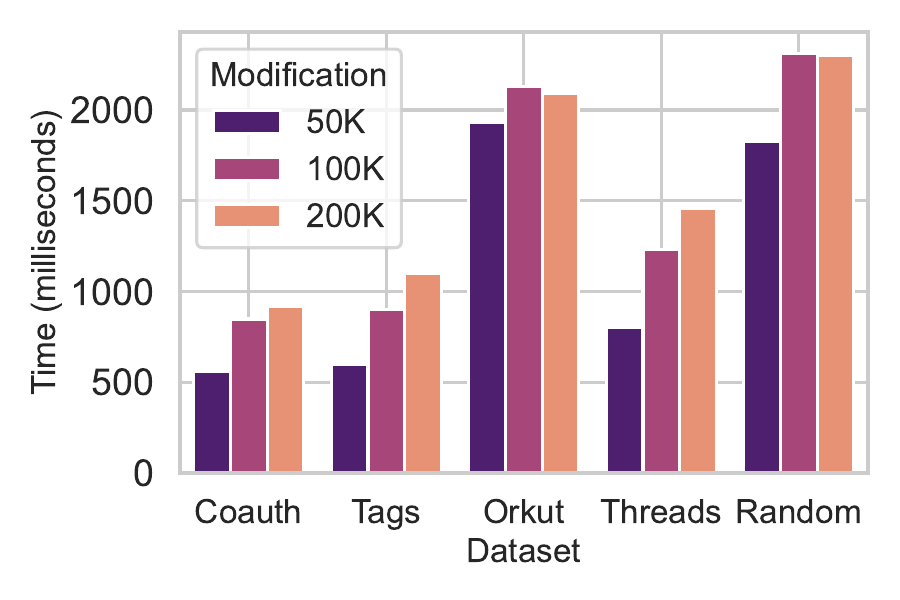}
	}

\caption{Effects of different hypergraph dynamics on triad counting execution time using \ourds.}
	\label{fig:5 datasets}
 \vspace{-0.2in}
 
\end{figure*}


\section{Performance evaluation}
\label{sec:performance}

We used an NVIDIA A100 GPU (80GB HBM2e) with a 64-core AMD EPYC Milan 7713 CPU (32GB memory) as host for our experiments. 
The datasets used are listed in Table~\ref{tab:my_label}.


\begin{table}
    \centering
    \small
    \caption{Dataset}
    
    \label{tab:my_label}
    \begin{tabular}{l|c| c | c}
    \hline
        Dataset & Hyperedges & Vertices &  Cardinality\\
        \hline
        \textbf{Coauth}~\cite{benson2018simplicial} & 2,599,087 & 1,924,991 & 280\\
        \textbf{Tags}~\cite{benson2018simplicial} & 5,675,497 & 49,998 & 4\\
        \textbf{Orkut}~\cite{leskovec2016snap} & 6,288,363 & 3,072,441 & 27K\\
        \textbf{Threads}~\cite{benson2018simplicial} & 9,705,709 & 2,675,955 & 67\\
        \textbf{Random} & 15,000,000 & 5,000,000 & 10000\\
        \hline
    \end{tabular}
    \vspace{-0.2in}
\end{table}

\subsection{Analysis of \ourds Operations in Triad Counting}

We test \ourds operations on dynamic hypergraphs for counting the hyperedge triads defined in MoCHy~\cite{lee2020hypergraph}.

\vspace{0.05in}
\textit{Hyperedge Modification:} We consider batches of changed hyperedges with 50\% deletions and 50\% insertions and compute the updated hyperedge triad counts. Fig.~\ref{fig:subfig1} shows the execution time for updating triad counts with batch sizes of $50K, 100K,$ and $200K$. The execution time increases with the batch size, but is not too large with respect to 
 the overall hypergraph size. This is because although the balanced binary search tree height grows with the dataset size, tree operations remain sublinear relative to the hypergraph growth.

We next increased the total hypergraph size and observed the execution time.
We generated random hypergraphs by varying the number of hyperedges from $20$ million to $55$ million, while setting the total number of vertices to one third of the hyperedges and a maximum cardinality of $10000$ per edge.
The number of changed hyperedges is fixed at $50$K.
Fig.~\ref{fig:varying_graph_size} shows that the execution time increases almost linearly to the number of hyperedges, indicating good scalability.


We study the effect of inserted hyperedge cardinality on hyperedge triad counting. Fig.~\ref{fig:subfig3} reports results for five datasets, each with $50K$ randomly generated hyperedges. The hyperedges  cardinalities are capped at $50$, $100$, and $200$ incident vertices per hyperedge. The results show  high cardinality insertions trigger more overflows that require allocating new memory blocks, as described in Section~\ref{subsec:ops_on_ds}, thereby increasing the time required to manage the data structure.

\vspace{0.05in}

\textit{Incident Vertex Modification:} Next, we study the impact of incident vertex modification on triad counting. Fig.~\ref{fig:subfig4} reports the execution time for updating counts with modified incident vertex batches of size $50K, 100K,$ and $200K$, where 50\% of the operations are insertions and 50\% are deletions. The results show that the Coauth, Tags, and Threads datasets yield lower execution times compared to Orkut and Random. This difference arises because the hyperedge length in Coauth, Tags, and Threads is small (max. 26), whereas in Orkut it is high (27k) and in Random it is fixed at 200.
The results reveal a positive correlation between the time required for incident vertex operations and the average hyperedge cardinality. 

\begin{figure}[!hbtp]
	\centering
	\subfloat[Coauth]{%
		\includegraphics[width=0.32\linewidth]{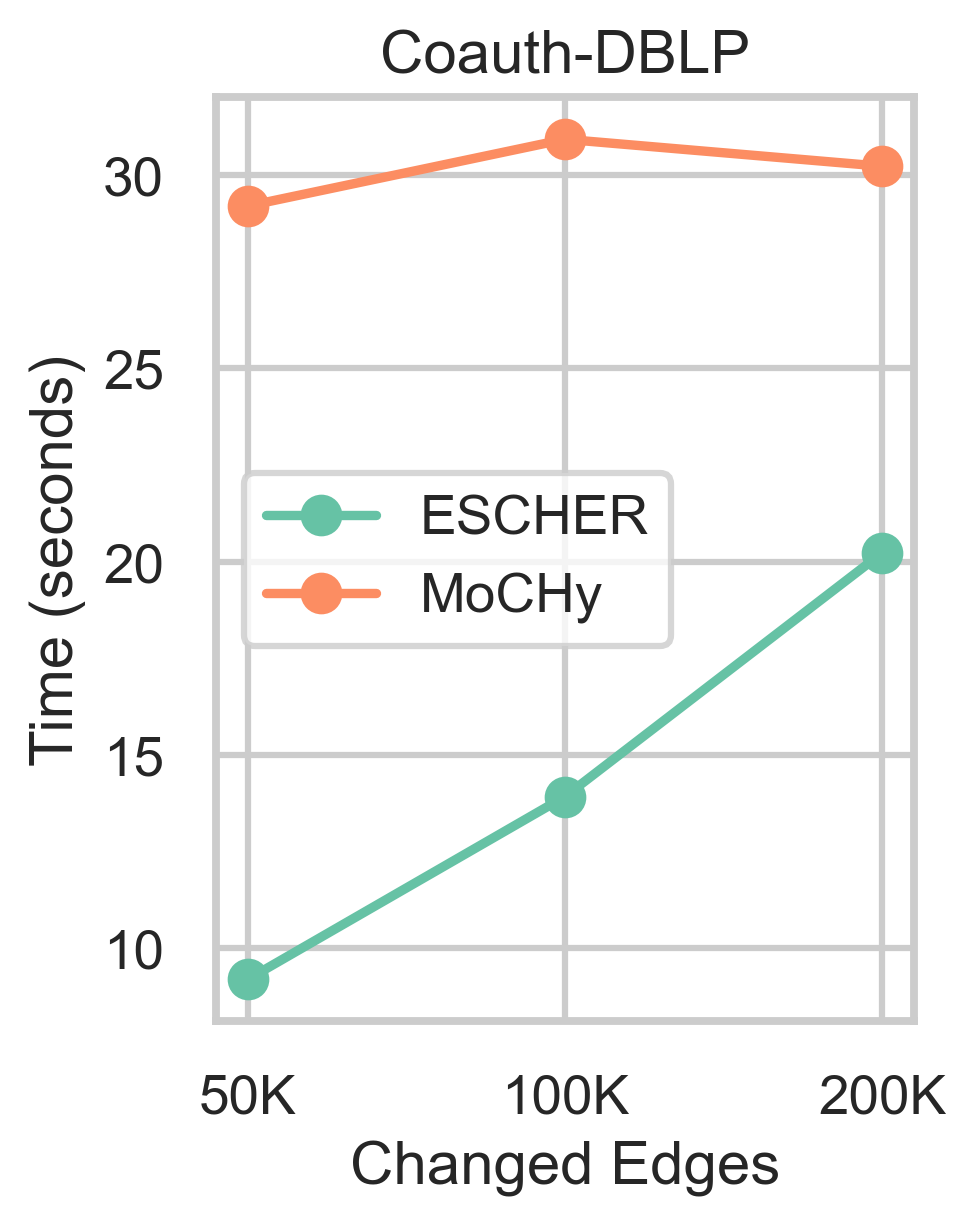}
	}
 \subfloat[Orkut]{%
		\includegraphics[width=0.33\linewidth]{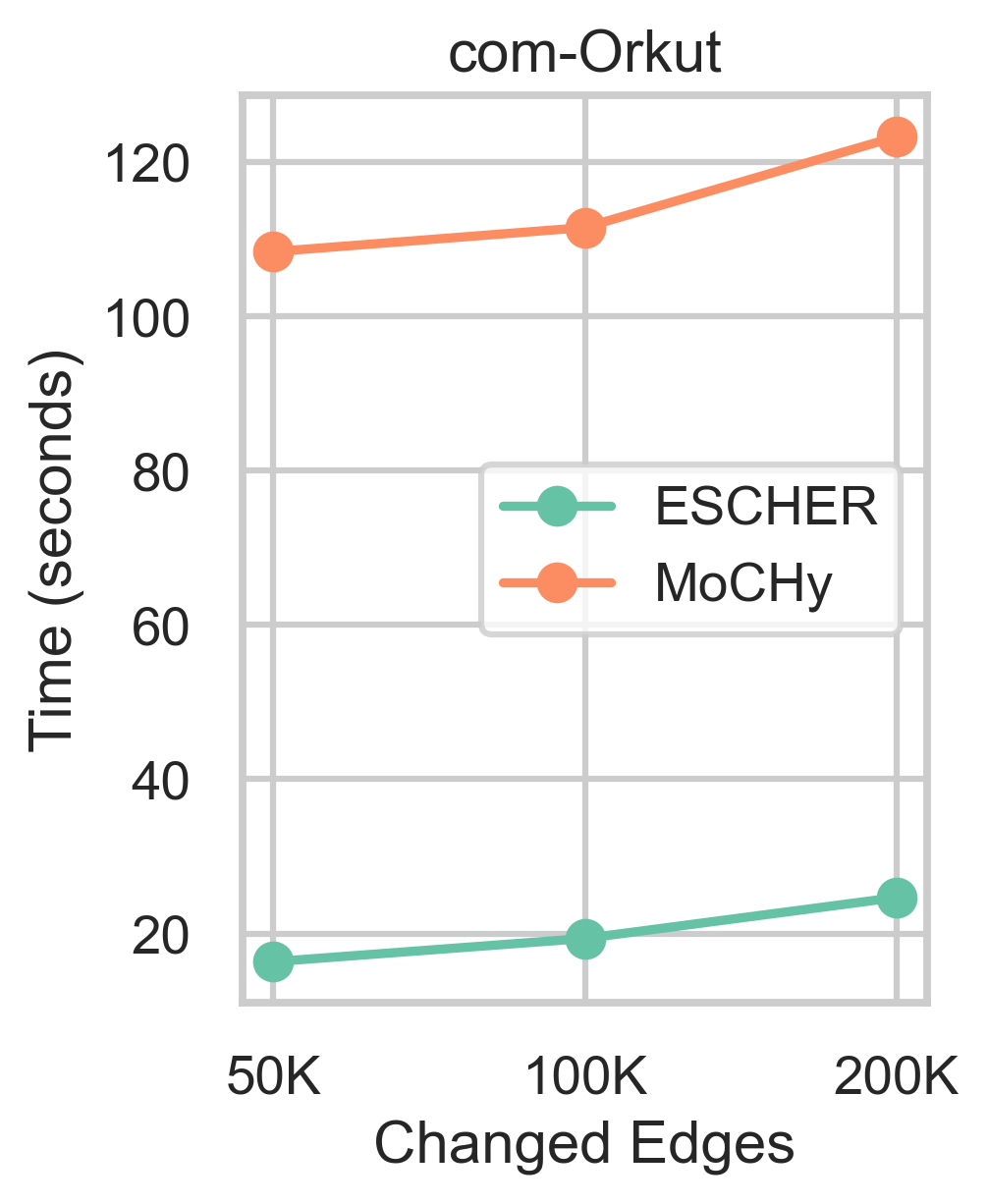}
	}
    \subfloat[Random]{%
		\includegraphics[width=0.33\linewidth]{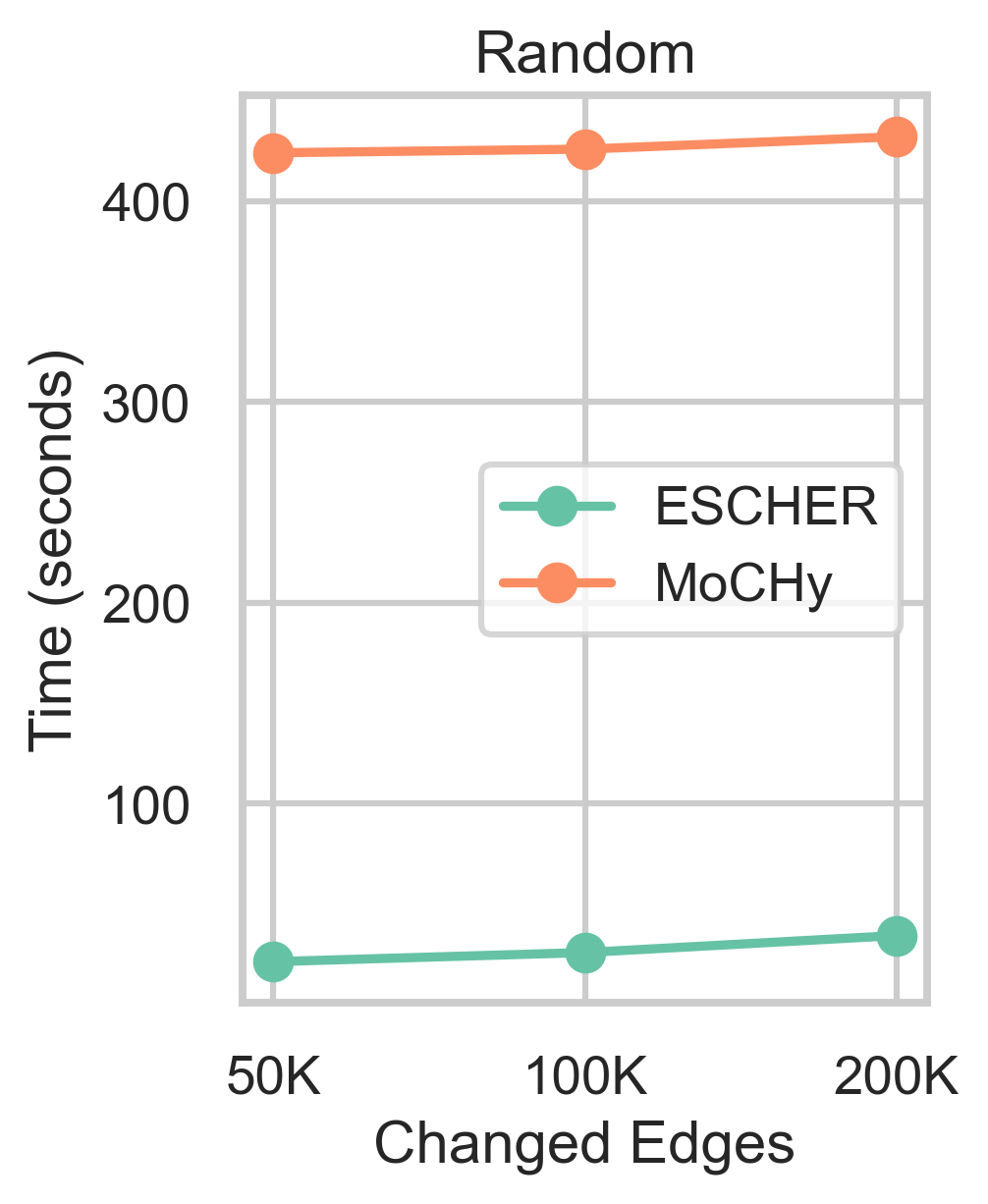}
	}
	\caption{Execution time analysis under varying hyperedge batch size.}
	\label{fig:staticvsdynamic}
 
\end{figure}

\begin{figure}[!hbtp]
	\centering
	\subfloat[Coauth]{%
		\includegraphics[width=0.315\linewidth]{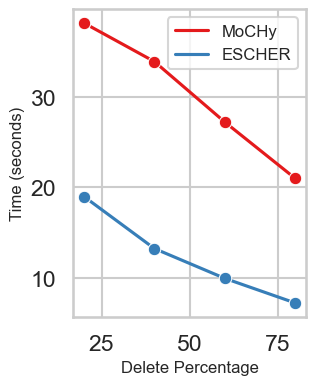}
	}
 \subfloat[Orkut]{%
		\includegraphics[width=0.33\linewidth]{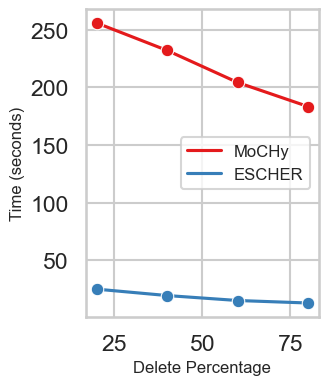}
	}
 \subfloat[Random]{%
	\includegraphics[width=0.33\linewidth]{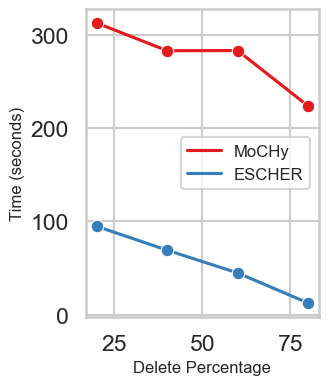}
	}
	\caption{Execution time analysis under varying deletion percentage. }
	\label{fig:delchange}
\vspace{-0.2in}
\end{figure}

\begin{figure}[!hbtp]
    \centering
    \begin{minipage}[b]{0.5\linewidth}
        \centering
        \includegraphics[width=\linewidth]{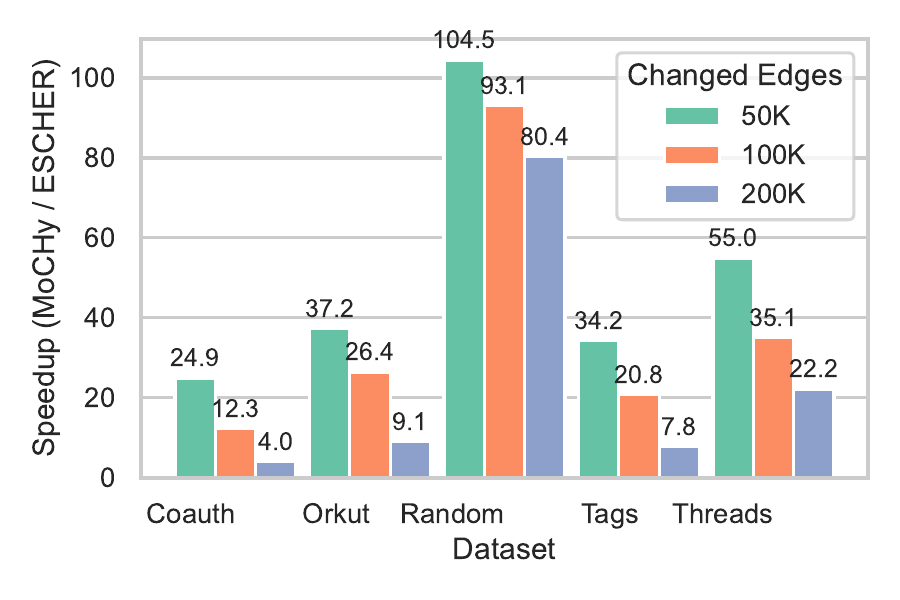}
        \caption{Comparison with \\MoCHy.}
        \label{fig:baseline}
    \end{minipage}
    \hspace{-0.15in}
    \begin{minipage}[b]{0.5\linewidth}
        \centering
        \includegraphics[width=\linewidth]{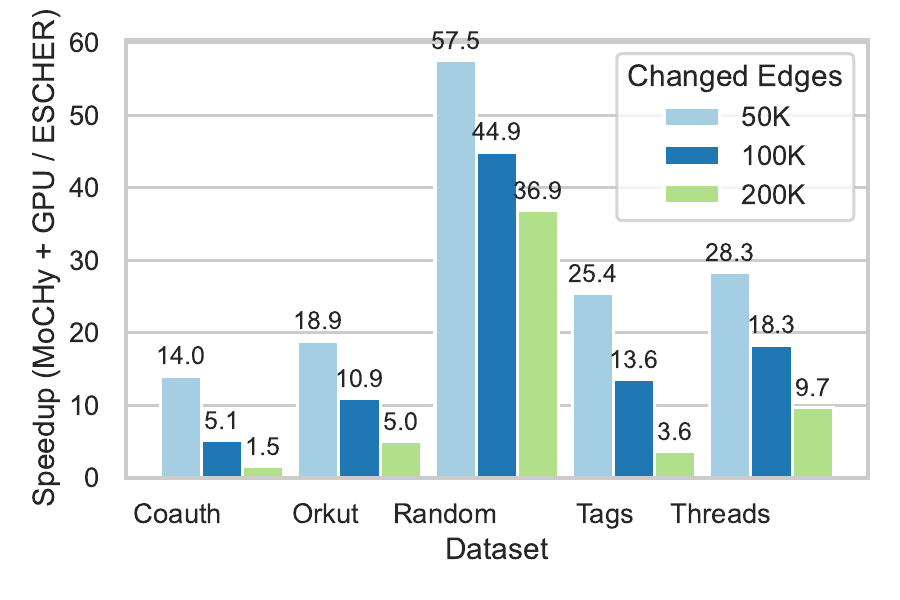}
        \caption{Comparison with GPU implementation of MoCHy }
        \label{fig:baselinecuda}
    \end{minipage}
\end{figure}

\subsection{Experiment on Hyperedge Triad Counting}
We compare the performance of dynamic hyperedge triad counting using \ourds with MoCHy~\cite{lee2020hypergraph}. Since MoCHy operates only on static hypergraphs, for each insertion or deletion batch, we first modify the hypergraph and then rerun MoCHy on the updated structure. Thus, MoCHy’s execution time excludes dynamic hypergraph maintenance. In contrast, the execution time of \ourds includes processing the updates, maintaining the hypergraph, and updating the triad counts.

Fig.~\ref{fig:staticvsdynamic} shows the execution time of both approaches as the total number of changed hyperedges per batch varies, with each batch containing $50\%$ insertions and $50\%$ deletions. When the number of changed hyperedges is small relative to the total hyperedges, \ourds significantly outperforms MoCHy. However, as the number of changes increases, the workload of \ourds grows while MoCHy’s remains constant.  Overall, \ourds achieves up to $19.32\times$ speedup with an average of $8.55\times$ over MoCHy.

Fig.~\ref{fig:delchange} shows the execution time as the percentage of hyperedge deletions varies. We tested deletion ratios of $20\%, 40\%, 60\%, 80\%$ in batches of $50K$ changes, where an $x\%$ deletion corresponds to $50000 \times x/100$ deletions and $50000 \times (100-x)/100$ insertions. On average, \ourds achieves a $6.5\times$ speedup over MoCHy, with a maximum of $24.37\times$. As the deletion percentage increases, the execution times of both static and dynamic versions decrease: for MoCHy, this is due to reduced recalculation overhead, while for \ourds it results from fewer hyperedge insertions, which are comparatively more expensive than deletions.

Fig.~\ref{fig:baseline} presents the speedup of hyperedge triad count update using \ourds compared to MoCHy's recomputation, under varying changed hyperedge batch sizes. The results show that speedup grows with dataset size but declines as the number of changed hyperedges increases. \ourds is on average $37.8 \times$, and up to $104.5 \times$ faster than MoCHy.

For fair comparison, we also implemented MoCHy with CUDA, and Fig.~\ref{fig:baselinecuda} reports the speedup of our approach over MoCHy's GPU version. Unlike \ourds, MoCHy's GPU implementation requires host–device back and forth data transfers to update the hypergraph. Even without accounting for the memory transfer overhead of MoCHy (GPU version), our triad count update method based on \ourds achieves an average speedup of $19.5\times$, reaching up to $57.5\times$.

\subsection{Experiment on Incident Vertex-based Triad Counting}

\begin{figure}[!hbtp]
	\centering
\vspace{-0.1in}	
\includegraphics[width=\linewidth]{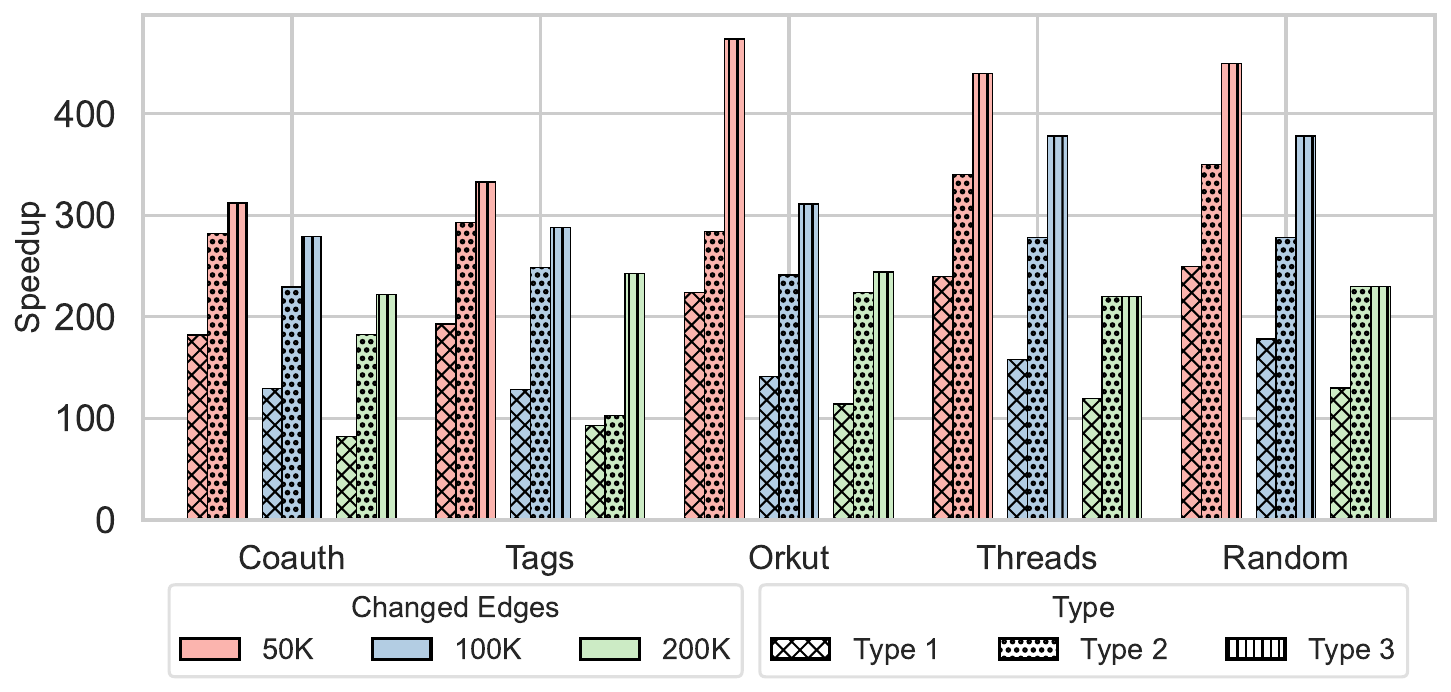}
        \caption{Execution time comparison with StatHyper}
        \label{fig:comparison_StatHyper}
\end{figure}

\begin{figure}[!hbtp]
	\centering
 
	\subfloat[\scriptsize  Execution time while varying the
number of hyperedge changes\label{fig:subfig1.1}]{%
		\includegraphics[width=0.45\linewidth]{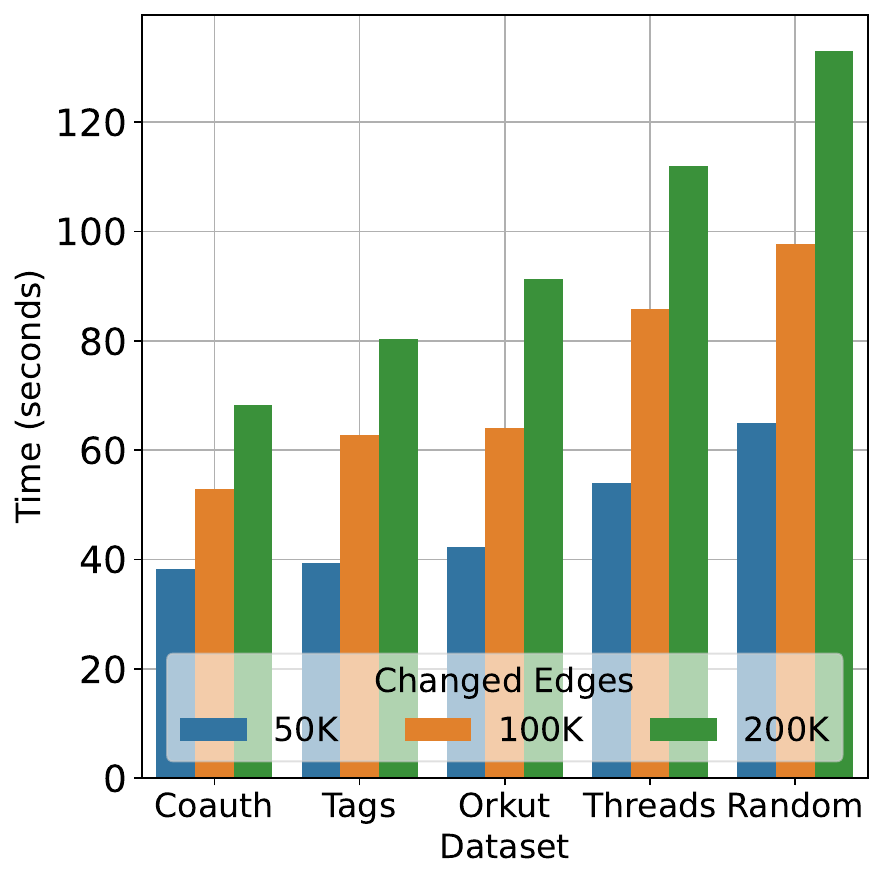}
	}
        \subfloat[\scriptsize Proportional Time taken by different steps \label{fig:subfig2.1}]{%
		\includegraphics[width=0.55
        \linewidth]{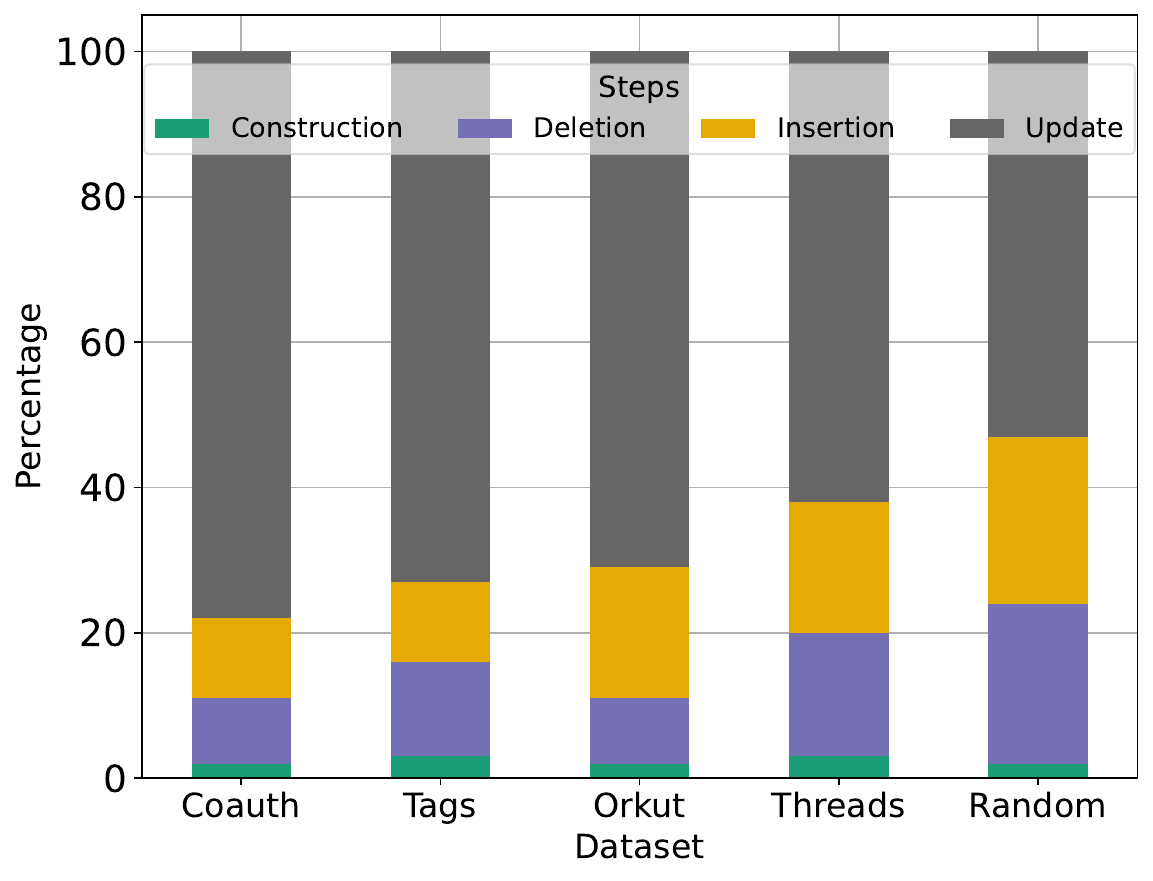}
	}
\caption{Experiments on five different datasets for temporal graph}
	\label{fig:5 datasetsTemporal}
 
\end{figure}

\begin{figure}[!hbtp]
	\centering
	\subfloat[Coauth]{%
		\includegraphics[width=0.33\linewidth]{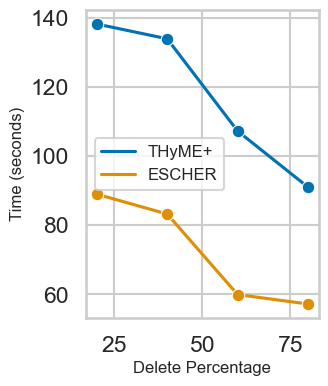}
	}
 \subfloat[Orkut]{%
		\includegraphics[width=0.33\linewidth]{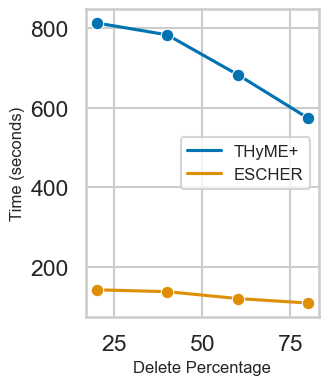}
	}
 \subfloat[Random]{%
	\includegraphics[width=0.33\linewidth]{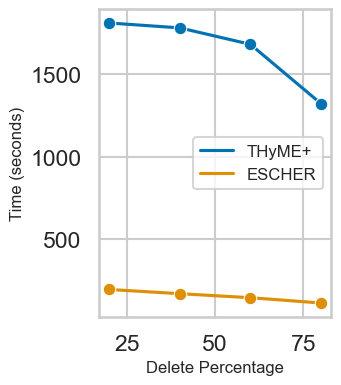}
	}
	\caption{\ourds vs. THyME+ comparison for different datasets varying delete percentage}
	\label{fig:temporalDatasets}
\vspace{-0.2in}
\end{figure}

\begin{figure}[!hbtp]
    \centering
    \begin{minipage}[b]{0.48\linewidth}
        \centering
        \includegraphics[width=\linewidth]{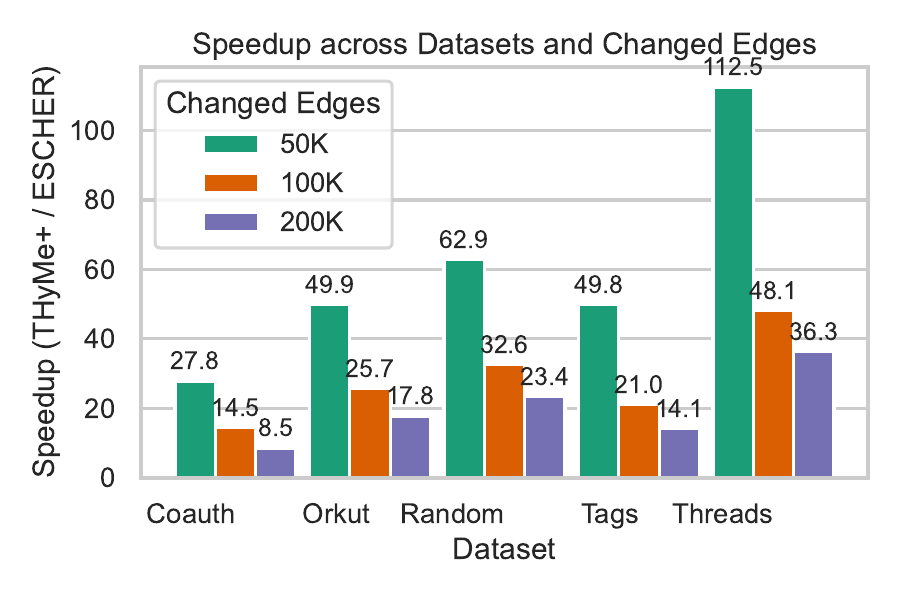}
        \caption{Comparison with THyMe+}
        \label{fig:baselineTHyMe}
    \end{minipage}
    \hfill
    \begin{minipage}[b]{0.48\linewidth}
        \centering
        \includegraphics[width=\linewidth]{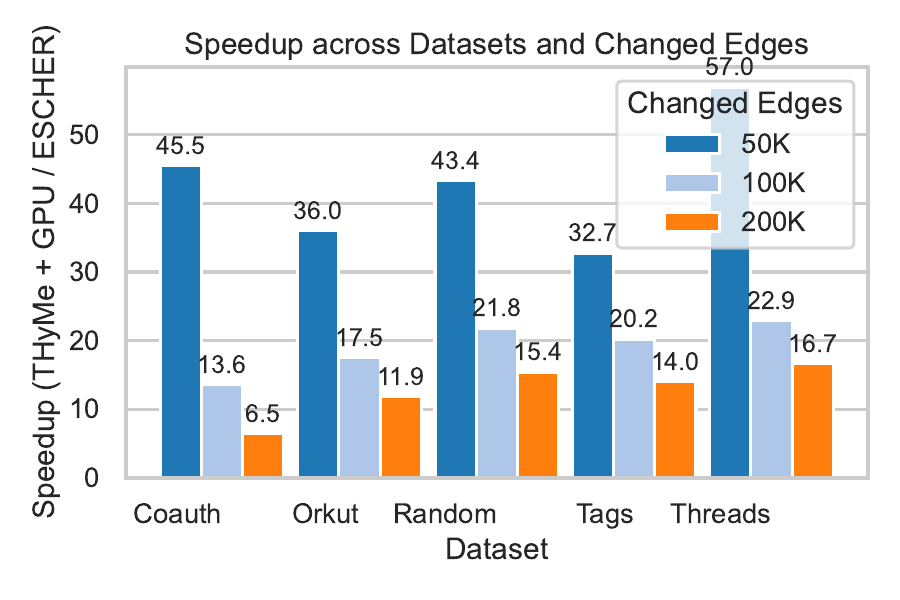}
        \caption{Comparison with THyMe+ GPU}
        \label{fig:baselinecudaTHyMe}
    \end{minipage}
\end{figure}

\begin{figure}[!hbtp]
	\centering
\vspace{-0.1in}	
		\includegraphics[width=0.6\linewidth]{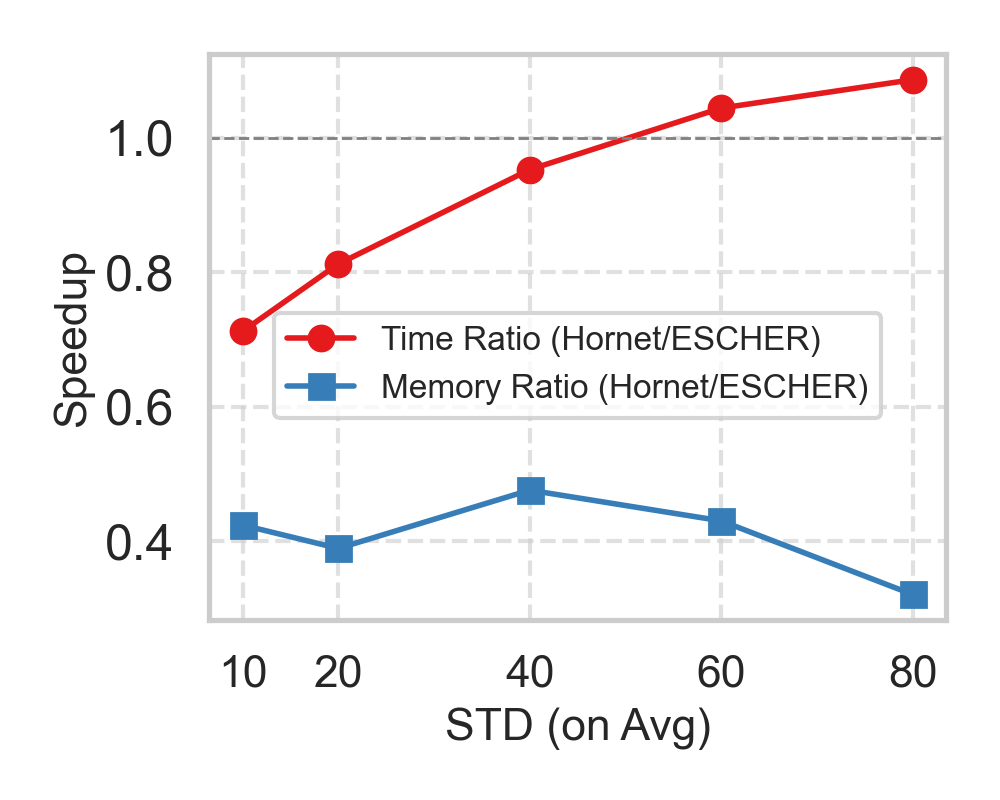}	
\caption{Execution time ratio between Hornet and \ourds, while varying changed edge sizes}
	\label{fig:typesp}
 \vspace{-0.2in}
 
\end{figure}

{
For the experiments on incident-vertex-based triad counting, we adopt triad types 1, 2, and 3 from StatHyper~\cite{bhattacharya2025statistical} and implement an incident-vertex-based triad count update method for dynamic hypergraphs as per Algorithm~\ref{algo:count_update}. As the original StatHyper is based on the R package \texttt{igraph}, it is not scalable. Therefore, we implement a CUDA-based StatHyper baseline that computes triad counts on static hypergraphs. 
We vary the size of the batch of changed hyperedges across datasets and, for each dynamic hypergraph snapshot, update triad counts using \ourds-based approach and recompute triad counts using StatHyper.

Fig.~\ref{fig:comparison_StatHyper} shows that, on average, for triad types 1, 2, and 3, \ourds achieves speedups of $157.4\times$, $252.1\times$, and $320.1\times$, with maximums of $249.8\times$, $349.8\times$, and $473.7\times$, respectively.
We observe that the speedup depends primarily on the size of the modified hyperedge batch; as the number of changed edges grows, the speedup decreases across all motif types. Across datasets and edge sizes, type 3 yields higher speedup than type 1. Recomputing type 3 triads is costlier by definition, requiring three incident vertices from three different hyperedges, and the static recomputation scans the whole snapshot to find them, whereas the update approach operates only on affected subgraphs. Consequently, recomputation slows more than the parallel update, increasing the relative speedup.

\subsection{Experiment on Temporal Triad Counting}
Fig.~\ref{fig:5 datasetsTemporal} reports experiments on updating temporal hypergraph motif counts in dynamic hypergraphs, restricting the temporal triad window to three consecutive timestamps. Fig.~\ref{fig:5 datasetsTemporal}(a) shows execution time when varying the number of changed edges (50\% insertion/deletion) per timestamp, generated by randomly selecting hyperedge IDs and their incident vertices. 
As the temporal triads are triads with time dependence among the hyperedges, the execution time pattern of updating temporal triads under various dynamics resembles the pattern observed in hyperedge-based triad count experiments.
While the overall trend of Fig.~\ref{fig:5 datasetsTemporal}(a) resembles Fig.~\ref{fig:5 datasets}(a), execution time is higher in temporal triads due to the overhead of maintaining and computing temporal dependencies. Fig.~\ref{fig:5 datasetsTemporal}(b) presents the proportion of time spent on data structure management (construction, deletion, insertion) and temporal triad count update.

We compare the \ourds-based temporal triad-count update approach with the state-of-the-art method THyMe+~\cite{lee2021thyme+}. Using a changed-hyperedge batch of size $50$K and varying deletion percentages, Fig.~\ref{fig:temporalDatasets} shows the execution time of temporal triad counting with \ourds and THyMe+.  The speedup in Fig.~\ref{fig:baselineTHyMe} shows that \ourds achieves up to $112.5\times$ speedup and $36.3\times$ on average across all datasets over the original shared-memory implementation of THyMe+. To ensure fairness, we also implemented THyMe+ on the GPU, and Fig.~\ref{fig:baselinecudaTHyMe} presents the corresponding speedups, where \ourds is up to $57\times$ faster and $25\times$ on average compared with GPU-based THyMe+.

\subsection{Comparison with Dynamic Graph Methods}
\color{blue}

\color{black}


Fig.~\ref{fig:typesp} compares \ourds  with Hornet~\cite{busato2018hornet}, a data structure for dynamic graphs with dyadic interactions. For fairness, we focus only on vertex-based triad type 1, as it can be represented in a dyadic interaction graph and thus allows direct comparison. On the X-axis, we vary the standard deviation (STD) of the cardinality in changed edges. 
The Y-axis shows the execution time ratio (Hornet divided by \ourds). For small STD values, \ourds is slower. For larger STD values, \ourds outperforms Hornet. 

This behavior reflects the memory management strategies: Hornet allocates memory in powers of two, which leads to costly memory copy operations when  STD is large.  \ourds preallocates memory in blocks that are multiples of 32 and manages them with linked lists, avoiding such copy overhead. However, for small STD values, \ourds incurs additional latency because it must traverse to the last entry of each block to either move to the next block or locate the end. 
Although for certain patterns of changed edges, \ourds can outperform Hornet, we observe that \ourds allocates more memory than Hornet to facilitate the storage of higher-order interactions.

}

\begin{table*}[!t]
\centering
\small
\caption{Speedup of \ourds compared to state-of-the-art hypergraph triad detection methods.}
\label{tab:speedup}
\begin{tabular}{|cc|cc|cc|cc|cc|}
\hline
\multicolumn{2}{|c|}{MoCHy (Shared mem.)} &
\multicolumn{2}{|c|}{MoCHy (GPU)} &
\multicolumn{2}{|c|}{THyMe+ (Shared mem.)} &
\multicolumn{2}{|c|}{THyMe+ (GPU)} &
\multicolumn{2}{|c|}{StatHyper (GPU)} \\
\hline
Avg & Max & Avg & Max & Avg & Max & Avg & Max & Avg & Max \\
\hline
$37.8\times$ & $104.5\times$ &
$19.5\times$ & $57.5\times$ &
$36.3\times$ & $112.5\times$ &
$25\times$ & $57\times$ &
$243.2\times$ & $473.7\times$ \\
\hline
\end{tabular}
\end{table*}

\subsection{Summary of Results}
Empirical analysis on four large real-world datasets and one large synthetic dataset shows that our triad count update approach using \ourds outperforms state-of-the-art techniques. A summary of speedups is listed in Table~\ref{tab:speedup}. 
To investigate the extended applicability of \ourds to dynamic graphs (dyadic interactions), we also compared with Hornet. However, Hornet supports only vertex-based triads. As \ourds incurs extra overhead for facilitating dynamic hypergraphs, Hornet generally outperforms \ourds on graphs. However, when the variability in the cardinalities of changed edges is high, \ourds shows better or comparable performance.

\section{Related work}
\label{related-work}

{\em Hypergraph Motif Counting.} MoCHy~\cite{lee2020hypergraph} defined 26 hypergraph triads based on hyperedge connectivity and proposed shared-memory parallel algorithms, both exact and heuristic, to count them. 
In contrast, \cite{niu2023size} emphasized that the relative sizes of hyperedge intersections within motifs provide richer information. By redefining hypergraph triads, they developed algorithms to identify top-$K$ triads optimizing intersection sizes. 
Beyond hyperedges, \cite{adler2022emergence} introduced hyper-motifs to study interactions among network motifs and uncover properties not present in individual motifs.

The authors in~\cite{bhattacharya2025statistical} study statistical inference for subgraph counts under an exchangeable hyperedge model, introducing three types of incident-vertex based hypergraph motifs that incorporate edge multiplicity. 
They also demonstrate that a subclass of these statistics remains robust to the deletion of low-degree nodes, enabling reliable inference in a dynamic hypergraph setting.
In~\cite{zhang2023efficiently}, the authors proposed a sample-and-estimate framework for hypergraph streams that achieves lower variance. THyMe+~\cite{lee2021thyme+} extended MoCHy~\cite{lee2020hypergraph} by defining 96 temporal hypergraph motifs to capture polyadic interaction patterns across consecutive timestamps. But it did not provide a parallel implementation. 
To address scalability, \cite{liu2019sampling} proposed sampling methods on temporal data, yielding memory-efficient and scalable algorithms.



\vspace{0.05in}
{\em Data Structure for Hypergraph Motif Counting.}
\label{datastruct}
The most feature-rich tool for hypergraph analysis, HGX~\cite{10.1093/comnet/cnad019}, lacks parallel processing and does not scale well to large networks. Parallelized tools such as HYGRA~\cite{10.1145/3332466.3374527} and CHGL~\cite{jenkins2018chapel} do not support dynamicity. HGX~\cite{10.1093/comnet/cnad019} and PAOH~\cite{8789484} provide visualization and diffusion simulation for dynamic hypergraphs, yet their support for  large-scale analysis remains limited. 

There exists several software for dynamic graphs. STINGER~\cite{ediger2013computational} employs an edge-list representation for update and analysis of large streaming and temporal graphs.  cuSTINGER~\cite{green2016custinger} adapts the GPU architecture to exploit parallelism. Hornet~\cite{busato2018hornet}  maintains a GPU-friendly memory layout for dynamic operations, often outperforming cuSTINGER. The work in~\cite{9767630} introduced the Loading-Processing-Switching (LPS) strategy to reduce redundant CPU–GPU transfers by loading only the necessary data partitions per batch, while~\cite{10058017} mitigated costly CSR rebalancing through the Leveled Packed Memory Array (LPMA), a binary tree-like data structure replacing the continuous array. For very large dynamic graphs that exceed DRAM capacity, persistent memory-based approaches have also been explored~\cite{sun2024ftgraph, wang2025scalable}.


\section{Conclusion}
\label{sec:conclusion}

We introduced the \ourds data structure to enable efficient and scalable dynamic hypergraph analysis. We also designed a framework for updating the counts of various categories of hypergraph triads. \ourds supports parallel hyperedge and incident-vertex insertions and deletions, ensuring high throughput on large datasets. Its effectiveness is demonstrated through extensive experiments on three categories of triad updates (hyperedge-based, incident-vertex-based, and temporal), where the \ourds-based framework consistently outperforms existing state-of-the-art methods by a significant margin. Our future work will focus on further optimizing memory utilization through more compact storage schemes, allowing even larger hypergraphs to be processed efficiently on the same hardware resources.




\bibliographystyle{IEEEtran}
\bibliography{sample-base}

@inproceedings{lee2021thyme+,
  title={Thyme+: Temporal hypergraph motifs and fast algorithms for exact counting},
  author={Lee, Geon and Shin, Kijung},
  booktitle={2021 IEEE International Conference on Data Mining (ICDM)},
  pages={310--319},
  year={2021},
  organization={IEEE}
}

@article{zhang2023efficiently,
  title={Efficiently Counting Triangles for Hypergraph Streams by Reservoir-Based Sampling},
  author={Zhang, Lingling and Zhang, Zhiwei and Wang, Guoren and Yuan, Ye and Zhao, Kangfei},
  journal={IEEE Transactions on Knowledge and Data Engineering},
  year={2023},
  publisher={IEEE}
}

@article{adler2022emergence,
  title={Emergence of dynamic properties in network hypermotifs},
  author={Adler, Miri and Medzhitov, Ruslan},
  journal={Proceedings of the National Academy of Sciences},
  volume={119},
  number={32},
  pages={e2204967119},
  year={2022},
  publisher={National Acad Sciences}
}

@article{niu2023size,
  title={Size-Aware Hypergraph Motifs},
  author={Niu, Jason and Amburg, Ilya D and Aksoy, Sinan G and Sar{\i}y{\"u}ce, Ahmet Erdem},
  journal={arXiv preprint arXiv:2311.07783},
  year={2023}
}

@inproceedings{liu2019sampling,
  title={Sampling methods for counting temporal motifs},
  author={Liu, Paul and Benson, Austin R and Charikar, Moses},
  booktitle={Proceedings of the twelfth ACM international conference on web search and data mining},
  pages={294--302},
  year={2019}
}

@article{ke2019community,
  title={Community detection for hypergraph networks via regularized tensor power iteration},
  author={Ke, Zheng Tracy and Shi, Feng and Xia, Dong},
  journal={arXiv preprint arXiv:1909.06503},
  year={2019}
}

@article{kong2019hypergraph,
  title={A hypergraph-based method for large-scale dynamic correlation study at the transcriptomic scale},
  author={Kong, Yunchuan and Yu, Tianwei},
  journal={BMC genomics},
  volume={20},
  pages={1--16},
  year={2019},
  publisher={Springer}
}

@inproceedings{wei2022dynamic,
  title={Dynamic hypergraph learning for collaborative filtering},
  author={Wei, Chunyu and Liang, Jian and Bai, Bing and Liu, Di},
  booktitle={Proceedings of the 31st ACM International Conference on Information \& Knowledge Management},
  pages={2108--2117},
  year={2022}
}

@inproceedings{green2016custinger,
  title={cuSTINGER: Supporting dynamic graph algorithms for GPUs},
  author={Green, Oded and Bader, David A},
  booktitle={2016 IEEE High Performance Extreme Computing Conference (HPEC)},
  pages={1--6},
  year={2016},
  organization={IEEE}
}

@inproceedings{busato2018hornet,
  title={Hornet: An efficient data structure for dynamic sparse graphs and matrices on gpus},
  author={Busato, Federico and Green, Oded and Bombieri, Nicola and Bader, David A},
  booktitle={2018 IEEE High Performance extreme Computing Conference (HPEC)},
  pages={1--7},
  year={2018},
  organization={IEEE}
}

@article{10.1093/comnet/cnad019,
    author = {Lotito, Quintino Francesco and Contisciani, Martina and De Bacco, Caterina and Di Gaetano, Leonardo and Gallo, Luca and Montresor, Alberto and Musciotto, Federico and Ruggeri, Nicolò and Battiston, Federico},
    title = "{Hypergraphx: a library for higher-order network analysis}",
    journal = {Journal of Complex Networks},
    volume = {11},
    number = {3},
    pages = {cnad019},
    year = {2023},
    month = {05},
    issn = {2051-1329},
    doi = {10.1093/comnet/cnad019},
    url = {https://doi.org/10.1093/comnet/cnad019},
    eprint = {https://academic.oup.com/comnet/article-pdf/11/3/cnad019/50461094/cnad019.pdf},
}

@inproceedings{10.1145/3332466.3374527,
author = {Shun, Julian},
title = {Practical Parallel Hypergraph Algorithms},
year = {2020},
isbn = {9781450368186},
publisher = {Association for Computing Machinery},
address = {New York, NY, USA},
url = {https://doi.org/10.1145/3332466.3374527},
doi = {10.1145/3332466.3374527},
booktitle = {Proceedings of the 25th ACM SIGPLAN Symposium on Principles and Practice of Parallel Programming},
pages = {232–249},
numpages = {18},
location = {San Diego, California},
series = {PPoPP '20}
}

@inproceedings{jenkins2018chapel,
  title={Chapel hypergraph library (chgl)},
  author={Jenkins, Louis and Bhuiyan, Tanveer and Harun, Sarah and Lightsey, Christopher and Mentgen, David and Aksoy, Sinan and Stavcnger, Timothy and Zalewski, Marcin and Medal, Hugh and Joslyn, Cliff},
  booktitle={2018 IEEE high performance extreme computing conference (HPEC)},
  pages={1--6},
  year={2018},
  organization={IEEE}
}

@ARTICLE{8789484,
  author={Valdivia, Paola and Buono, Paolo and Plaisant, Catherine and Dufournaud, Nicole and Fekete, Jean-Daniel},
  journal={IEEE Transactions on Visualization and Computer Graphics}, 
  title={Analyzing Dynamic Hypergraphs with Parallel Aggregated Ordered Hypergraph Visualization}, 
  year={2021},
  volume={27},
  number={1},
  pages={1-13},
  doi={10.1109/TVCG.2019.2933196}}

@article{ediger2013computational,
  title={Computational graph analytics for massive streaming data},
  author={Ediger, David and Riedy, Jason and Bader, David A and Meyerhenke, Henning},
  journal={Large Scale Network-Centric Distributed Systems},
  pages={619--648},
  year={2013},
  publisher={Wiley Online Library}
}

@article{park2001parallel,
  title={Parallel algorithms for red--black trees},
  author={Park, Heejin and Park, Kunsoo},
  journal={Theoretical Computer Science},
  volume={262},
  number={1-2},
  pages={415--435},
  year={2001},
  publisher={Elsevier}
}

@inproceedings{khanda2023parallel,
  title={A Parallel Algorithm for Updating a Multi-objective Shortest Path in Large Dynamic Networks},
  author={Khanda, Arindam and Shovan, SM and Das, Sajal K},
  booktitle={Proceedings of the SC'23 Workshops of The International Conference on High Performance Computing, Network, Storage, and Analysis},
  pages={739--746},
  year={2023}
}

@inproceedings{haryan2022shared,
  title={Shared-memory parallel algorithms for fully dynamic maintenance of 2-connected components},
  author={Haryan, Chirayu Anant and Ramakrishna, G and Kothapalli, Kishore and Banerjee, Dip Sankar},
  booktitle={2022 IEEE International Parallel and Distributed Processing Symposium (IPDPS)},
  pages={1195--1205},
  year={2022},
  organization={IEEE}
}

@inproceedings{sahu2024shared,
  title={Shared-Memory Parallel Algorithms for Community Detection in Dynamic Graphs},
  author={Sahu, Subhajit and Kothapalli, Kishore and Banerjee, Dip Sankar},
  booktitle={2024 IEEE International Parallel and Distributed Processing Symposium Workshops (IPDPSW)},
  pages={250--259},
  year={2024},
  organization={IEEE}
}

@inproceedings{ccatalyurek2001fine,
  title={A Fine-Grain Hypergraph Model for 2D Decomposition of Sparse Matrices.},
  author={{\c{C}}ataly{\"u}rek, {\"U}mit V and Aykanat, Cevdet},
  booktitle={IPDPS},
  volume={1},
  pages={118},
  year={2001},
  organization={Citeseer}
}

@inproceedings{liu2022high,
  title={High-order line graphs of non-uniform hypergraphs: Algorithms, applications, and experimental analysis},
  author={Liu, Xu T and Firoz, Jesun and Aksoy, Sinan and Amburg, Ilya and Lumsdaine, Andrew and Joslyn, Cliff and Praggastis, Brenda and Gebremedhin, Assefaw H},
  booktitle={2022 IEEE International Parallel and Distributed Processing Symposium (IPDPS)},
  pages={784--794},
  year={2022},
  organization={IEEE}
}

@article{benson2018simplicial,
  title={Simplicial closure and higher-order link prediction},
  author={Benson, Austin R and Abebe, Rediet and Schaub, Michael T and Jadbabaie, Ali and Kleinberg, Jon},
  journal={Proceedings of the National Academy of Sciences},
  volume={115},
  number={48},
  pages={E11221--E11230},
  year={2018},
  publisher={National Acad Sciences}
}

@article{leskovec2016snap,
  title={Snap: A general-purpose network analysis and graph-mining library},
  author={Leskovec, Jure and Sosi{\v{c}}, Rok},
  journal={ACM Transactions on Intelligent Systems and Technology (TIST)},
  volume={8},
  number={1},
  pages={1--20},
  year={2016},
  publisher={ACM New York, NY, USA}
}

@incollection{bell2012thrust,
  title={Thrust: A productivity-oriented library for CUDA},
  author={Bell, Nathan and Hoberock, Jared},
  booktitle={GPU computing gems Jade edition},
  pages={359--371},
  year={2012},
  publisher={Elsevier}
}

@inproceedings{sun2024ftgraph,
  title={FTGraph: A Flexible Tree-Based Graph Store on Persistent Memory for Large-Scale Dynamic Graphs},
  author={Sun, Gan and Zhou, Jiang and Li, Bo and Gu, Xiaoyan and Wang, Weiping and He, Shuibing},
  booktitle={2024 IEEE International Conference on Cluster Computing (CLUSTER)},
  pages={39--50},
  year={2024},
  organization={IEEE}
}

@article{wang2025scalable,
  title={Scalable and High-Performance Large-Scale Dynamic Graph Storage and Processing System},
  author={Wang, Rui and Zong, Weixu and He, Shuibing and Li, Yongkun and Xu, Yinlong},
  journal={ACM Transactions on Storage},
  year={2025},
  publisher={ACM New York, NY}
}

@ARTICLE{10058017,
  author={Zou, Lei and Zhang, Fan and Lin, Yinnian and Yu, Yanpeng},
  journal={IEEE Transactions on Knowledge and Data Engineering}, 
  title={An Efficient Data Structure for Dynamic Graph on GPUs}, 
  year={2023},
  volume={35},
  number={11},
  pages={11051-11066},
  doi={10.1109/TKDE.2023.3235941}}

@ARTICLE{9767630,
  author={Zhang, Yu and Liang, Yuxuan and Zhao, Jin and Mao, Fubing and Gu, Lin and Liao, Xiaofei and Jin, Hai and Liu, Haikun and Guo, Song and Zeng, Yangqing and Hu, Hang and Li, Chen and Zhang, Ji and Wang, Biao},
  journal={IEEE Transactions on Knowledge and Data Engineering}, 
  title={EGraph: Efficient Concurrent GPU-Based Dynamic Graph Processing}, 
  year={2023},
  volume={35},
  number={6},
  pages={5823-5836},
  doi={10.1109/TKDE.2022.3171588}}

@article{shovan2025parallel,
  title={Parallel Multi Objective Shortest Path Update Algorithm in Large Dynamic Networks},
  author={Shovan, SM and Khanda, Arindam and Das, Sajal K},
  journal={IEEE Transactions on Parallel and Distributed Systems},
  year={2025},
  publisher={IEEE}
}

@article{lee2020hypergraph,
author = {Lee, Geon and Ko, Jihoon and Shin, Kijung},
title = {Hypergraph motifs: concepts, algorithms, and discoveries},
year = {2020},
issue_date = {August 2020},
publisher = {VLDB Endowment},
volume = {13},
number = {12},
issn = {2150-8097},
url = {https://doi.org/10.14778/3407790.3407823},
doi = {10.14778/3407790.3407823},
journal = {Proc. VLDB Endow.},
month = jul,
pages = {2256–2269},
numpages = {14}
}

@article{bhattacharya2025statistical,
  title={Statistical Inference for Subgraph Frequencies of Exchangeable Hyperedge Models},
  author={Bhattacharya, Ayoushman and Chakraborty, Nilanjan and Lunde, Robert},
  journal={arXiv preprint arXiv:2508.13258},
  year={2025}
}

@inproceedings{fox2018fast,
  title={Fast and adaptive list intersections on the gpu},
  author={Fox, J. and Green, O. and Gabert, K. and An, X. and Bader, D. A.},
  booktitle={2018 IEEE High Performance extreme Computing Conference (HPEC)},
  pages={1--7},
  year={2018},
  organization={IEEE}
}

@inproceedings{green2014fast,
  title={Fast triangle counting on the GPU},
  author={Green, Oded and Yalamanchili, Pavan and Mungu{\'\i}a, Llu{\'\i}s-Miquel},
  booktitle={2014 4th Workshop on Irregular Applications: Architectures and Algorithms (IA\^{} 3)},
  pages={1--8},
  year={2014},
  organization={IEEE}
}

\end{document}